\documentclass[]{rsos}%%%%where rsos is the template name

\usepackage[super,sort&compress,comma]{natbib}

\begin{document}

%%%% Article title to be placed here
\title{Dust evolution, a global view: II. Top-down branching, nano-particle fragmentation \\ 
   and the mystery of the diffuse interstellar band carriers}

\author{%%%% Author details
A. P. Jones} %, X. Second author$^{2}$ and X. Third author$^{3}$}

%%%%%%%%% Insert author address here
\address{Institut d'Astrophysique Spatiale, CNRS, Univ. Paris-Sud, Universit\'e Paris-Saclay, B\^at. 121, 91405 Orsay cedex, France}

%%%% Subject entries to be placed here %%%%
\subject{Astrochemistry, Astrophysics}

%%%% Keyword entries to be placed here %%%%

%%%% Insert corresponding author and its email address}
\corres{A.P. Jones\\
\email{Anthony.Jones@ias.u-psud.fr}}

%%%% Abstract text to be placed here %%%%%%%%%%%%
\begin{abstract}
The origin of the diffuse interstellar bands is one of the longest-standing mysteries of the interstellar medium is explored within the framework of {\em The Heterogeneous dust Evolution Model at the IaS} (THEMIS). The likely nature of the diffuse interstellar band carriers and their evolution is here explored within the framework of the structures and sub-structures inherent to doped hydrogenated amorphous carbon grains in the interstellar medium. Based on the natural aromatic-rich moieties (ashphaltenes) recovered from coal and oil the likely structure of their interstellar analogues is investigated within the context of the diffuse band problem. It is here proposed that the top-down evolution of interstellar carbonaceous grains, and in particular a-C(:H) nano-particles, is at the heart of the formation and evolution of the diffuse interstellar band carriers and their associations with small molecules and radicals such as C$_2$, C$_3$, CH and CN. It is most likely that the diffuse interstellar bands are carried by dehydrogenated, ionised, hetero-cyclic, aromatic rich moieties that form an integral part of the contiguous structure of hetero-atom doped hydrogenated amorphous carbon nano-particles and their daughter fragmentation products.
\end{abstract}
%%%%%%%%%%%%%%%%%%%%%%%%%%%

\maketitle

%------------------------------------------------------------------
\section{Introduction}
%------------------------------------------------------------------

The search for the origin of the diffuse interstellar bands (DIBs) has been a long journey \cite[{\it e.g.},][]{1995ARA&A..33...19H,1995_DIBs_book,2014IAUS..297}; a journey that is far from over despite the recent assignment of two DIBs to the fullerene cation, C$_{60}^+$.\citep{2015Natur.523..322C}  

Here we explore the fundamental nature and the very roots of the likely DIB carriers, and their formation, by profiting from the most recent atomic resolution analyses of aromatic moieties. \citep{Gross_etal_2009,Schuler_etal_2014,2015NatCh...7..623P,Schuler_etal_2016} and naturally-occurring carbonaceous materials.\citep{JACS_2015_Asphaltenes} This work is also based on experimental and theoretical work on the the dehydrogenation of polycyclic aromatic hydrocarbons (PAHs) to form cluster cations C$_n$H$_x^+$ with $x \geq 0$. \citep[{\it e.g.},][]{1996A&A...305..602A,1996A&A...305..616A,
double_ring_dehydPAH2,double_ring_dehydPAH1,1999IJMSp.185....1G} 

The approach adopted here is generic\footnote{The ideas presented here and in the accompanying papers\cite{ANT_RSOS_nanoparticles,ANT_RSOS_globules} arise from a re-consideration of the nature of interstellar dust within the framework of {\em The Heterogeneous dust Evolution Model at the IaS} (THEMIS).\cite{2016A&A...999A..99J} A brief introductory description to THEMIS is given in Paper I\cite{ANT_RSOS_nanoparticles}.} and will therefore certainly not furnish the specific identification of a single DIB carrier. However, if this exploration can bring something to bear on the likely nature and evolution of solid carbonaceous matter and its relationship to the DIBs, especially at nano-particle and smaller size-scales, then it could lead to the elucidation of whole families of related but structurally-distinct species or grain sub-structures that may be central to an explanation for the DIBs, their origin, evolution and associations with particular molecules, ions and radicals. 

This paper is structured as follows:  
Section \ref{sect_riches} considers the challenge presented by the overwhelming variety of possible chemical structures, 
Section \ref{sect_character} proposes a new nomenclature scheme for simply characterising hetero-aromatic moieties, Section \ref{sect_dehyd} summarises some of the known effects of dehydrogenation and ionisation on aromatic species, 
Section \ref{sect_new_form} suggests a new form of hydrocarbon ring molecule,  
Section \ref{sect_ISM} investigates the likely nature of the DIB carriers in the interstellar medium (ISM), 
Section \ref{sect_topdown0} proposes a new top down branching route for nano-particle evolution and the formation of the DIB carriers in the ISM, 
Section \ref{sect_experiment} suggests possible follow-on investigations and 
Section \ref{sect_conclusions} concludes.

%------------------------------------------------------------------
\section{An embarrassment of riches?} 
\label{sect_riches}
%------------------------------------------------------------------ 

For now, if we consider only hydrocarbon molecules, ions or radicals, with perhaps single O or N hetero-atoms ({\it i.e.}, C$_n$H$_x$O or C$_n$H$_x$N species) we can simply characterise their contiguous structures by the number of multiply-bonded ($\geq 2$), heavy atoms (C, O or N), which make up their chain and/or ring structures. 
For simple species  containing up to six heavy atoms, {\it i.e.}, C$_6$H$_x$, C$_5$H$_x$O or C$_5$H$_x$N with $x \leq (2n+2)$, Table~\ref{table_variety} shows an estimate of the number of structurally-distinct species. 
The numbers in this table show the wide variety of possible chemical structures if we include only single O or N hetero-atoms, simple (de-)hydrogenation/saturation, neutral and singly ionised species.
Given that fully hydrogenated species are less likely to be ionised in denser regions, and ionised species are more likely to be dehydrogenated in ionised regions, we have conservatively divided the total number of possible species by two in order to try to estimate how many different configurations are likely to be of relevance to ISM chemistry. 
As Table~\ref{table_variety} shows, species with only six heavy atoms are likely to exhibit well over one thousand different structural forms. 
 The number of possible conformations probably goes with, at least, the fourth power of the number of atoms. Hence, species with about 30 heavy atoms will likely have more than a million possible structures. In any case, such small structures are probably not of significance because they will rapidly be destroyed  in the diffuse ISM by ultraviolet and extreme ultraviolet (UV and EUV) photons \citep{1994ApJ...420..307J,1996A&A...305..602A,1996A&A...305..616A} (hereafter referred to as UV) and only species with more that 50 carbon atoms are likely to survive.\citep{1996A&A...305..602A,1996A&A...305..616A}  
Clearly, for a bottom-up chemistry investigation, as the number of atoms increases a detailed study of all of the possible structures becomes intractable.

% TABLE 1
\begin{table}[!h]
\caption{Hydrocarbon chemical complexity in the ISM.}
\label{table_variety}
\begin{center}
\begin{tabular}{lccccc}
\hline
                                                 &                   &                 &                           &                                 &                                  \\[-0.25cm]
          Structure                        &   No. of      &   No. of   &   No. of                &   Minimum no. of     &    $\approx$ No. of    \\
          (  $p = 0$ or 1 )             &   atoms       &   isomers   &   hetero-atomic   &   (de-)hydrogenated  &    possible                 \\
         \{ charge states, $i=2$ \}   &     ( $n$ )  &     ( $b$ )   &  isomers  ( $h$ ) &    states  $(2n+3)$   &     forms                    \\    
                                                  &                  &                 &   [ cyclic form only ]  &                            &                                  \\
                                                &                   &                 &                               &                               &                                  \\[-0.3cm]
\hline
          &      &      &       &        \\[-0.25cm]  
       \ \ \ (CX$_p$) \ \,      \{ 0 / + \}   &    1   &  1   & 1 \ \ \ \ [ $0$ ]  &    5     &       5         \\[0.05cm]
       \ \ \ (CX$_p$)$_2$ \ \{ 0 / + \}   &    2   &  1   & 1 \ \ \ \ [ $0$ ]  &    7     &      14         \\[0.05cm]
          c-(CX$_p$)$_3$ \ \{ 0 / + \}   &    3   &  2   & 2 \ \ \ \ [ $2$ ]  &    9     &    108         \\[0.05cm]
          c-(CX$_p$)$_4$ \ \{ 0 / + \}   &    4   &  2   & 2 \ \ \ \ [ $0$ ]  &   11    &    176         \\[0.05cm]
          c-(CX$_p$)$_5$ \ \{ 0 / + \}   &    5   &  4   & 3 \ \ \ \ [ $1$ ]  &   13    &    780         \\[0.05cm]
          c-(CX$_p$)$_6$ \ \{ 0 / + \}   &    6   &  6   & 3 \ \ \ \ [ $1$ ]  &   15    &  1620         \\[0.05cm]
\hline
        &       &      &       &        &           \\[-0.25cm]
\end{tabular}  
\begin{list}{}{}
Notes:
\item[] 1. In column 1 the prefix c- indicates that a stable cyclic form exists {\em in addition} to the chain and branched-chain isomers. 
\item[] 2. The approximation of the number of most likely different forms is given by  \\ $(i \times n \times b \times h \times [2n+3] )/2$
\item[] 3. The dehydrogenated states (column 5) does not count the number of different configurations possible for each $n$ and is therefore only lower limit. 
\end{list}    
\end{center}
\end{table}

%------------------------------------------------------------------
\section{Polycyclic (hetero-)aromatic structure characterisation and nomenclature} 
\label{sect_character}
%------------------------------------------------------------------ 

Here we explore and characterise the basic carbonaceous material (sub-)structures occurring in natural materials (petroleum and coal asphaltenes) with the view that these same types of structures ought also to be present in carbonaceous grains in the ISM.

%------------------------------------------------------------------
\subsection{Structure nomenclature} 
\label{sect_nomenclature}
%------------------------------------------------------------------ 

In order to characterise carbonaceous frameworks a new nomenclature scheme is proposed here to enable a general description of asphaltenes-type structures and in order to provide a descriptor for the likely sub-structures in interstellar carbonaceous grains. Specifically, for the intrinsic five, six and seven-membered rings we adopt the following labels: \\
\hspace*{0.5cm} P = pentagonal, \, 5-fold rings \\
\hspace*{0.5cm} S = hexagonal, \ \, 6-fold rings \\
\hspace*{0.5cm} G = heptagonal, \,7-fold rings. \\
To enable the maximum flexibility in the use of aromatic structure descriptor the following rules and indicators are proposed: 
\begin{itemize}
\item The naming sequence begins with the largest S aromatic domain, unless this is a central moiety. 
\item A subscript $n$, {\it e.g.}, S$_n$, indicates the total number of rings ($>1$) of the given type within any distinct domain but does not consider the particular arrangement of the rings. 
\item Distinct ring domain names are concatenated, {\it e.g.}, S$_n$P$_m$, in the same arrangement as they appear in the structure. 
\item A superscript $^\prime$ indicates a methylene ($-$CH$_2-$) bridge in the given ring, {\it e.g.}, P$^\prime$ or S$^\prime$.
\item A superscript $^{\rm X}$ indicates a hetero-cycle containing one (two) hetero-atoms, {\it e.g.}, P$^{\rm X}$ (P$^{\rm XX}$), where X = N, O, S, Si, P, B, \ldots
\item Pendant S or P rings attached to the larger structure by a single bond are indicated by $-$S or $-$P, respectively. 
\end{itemize}

Table~\ref{table_nomen} shows some simple examples of how this nomenclature scheme can be applied to some simple aromatic and polycyclic aromatic species. 

% TABLE 2
\begin{table}
\caption{Polycyclic (hetero-)aromatic nomenclature.}
\begin{center}
\begin{tabular}{lll}
\hline
                                        &                                   &                     \\[-0.25cm]
         Name                     &   Formula                  & Descriptor    \\
                                        &                                  &                     \\[-0.3cm]
\hline  
                                       &                                  &                              \\[-0.25cm]
       benzene                  &   C$_6$H$_{6}$        &     S                      \\[0.05cm]
       naphthalene            &   C$_{10}$H$_{8}$    &    S$_2$                          \\[0.05cm]
       azulene                    &   C$_{10}$H$_{8}$   &    GP                              \\[0.05cm]
       biphenyl                   &   C$_{12}$H$_{10}$  &    S$-$S                        \\[0.05cm]
       fluorene                   &   C$_{13}$H$_{10}$  &    SP$^\prime$S           \\[0.05cm]
       anthracene              &   C$_{14}$H$_{10}$  &    S$_3$                          \\[0.05cm]
       phenanthrene          &   C$_{14}$H$_{10}$  &    S$_3$                          \\[0.05cm]
       pyrene                     &   C$_{16}$H$_{10}$  &    S$_4$                          \\[0.05cm]
       2,3-benzofluorene   &   C$_{17}$H$_{12}$  &    S$_2$P$^\prime$S     \\[0.05cm]
       benz[a]anthracene   &   C$_{18}$H$_{12}$  &    S$_4$                        \\[0.05cm]
       chrysene                  &   C$_{18}$H$_{12}$  &    S$_4$                        \\[0.05cm]
       triphenylene             &   C$_{18}$H$_{12}$  &    S$_4$                        \\[0.05cm]
       corannulene            &   C$_{20}$H$_{10}$  &    S$_5$P                       \\[0.05cm]
       perylene                  &   C$_{20}$H$_{12}$  &    S$_5$                         \\[0.05cm]
       benzo[ghi]perylene  &   C$_{22}$H$_{12}$  &    S$_6$                         \\[0.05cm]
       coronene                &   C$_{24}$H$_{12}$  &    S$_7$                         \\[0.05cm]
       fulllerene$^1$         &   C$_{60}$                 &    S$_{20}$12$\times$P   \\[0.05cm]
\hline
                                &                   &                     \\[-0.25cm]
\end{tabular}  
\begin{list}{}{}
Notes:
\item[] 1. In fullerene the 12 pentagons are isolated, hence, the descriptor should be S$_{20}$PPPPPPPPPPPP but this is rather cumbersome and so a simplified 12$\times$P descriptor is used above. 
\end{list}    
\end{center}    
  \label{table_nomen}
\end{table}

%------------------------------------------------------------------
\subsection{Asphaltenes: a framework guide} 
\label{sect_asphaltenes}
%------------------------------------------------------------------ 

Here we explore the structure of asphaltenes, which are natural, aromatic-rich species derived from petroleum (PA) and coal (CA). More than one hundred of these PA and CA asphaltene species were recently analysed by atomic force microscopy (AFM) and molecular orbital imaging using scanning tunnelling microscopy at atomic-resolution.\citep{JACS_2015_Asphaltenes} In the following we use the results of this study \cite{JACS_2015_Asphaltenes} as a basis for the discussion presented in this section.

Both PA and CA asphaltenes show a wide range of complex aromatic structures, often with attached methyl and large alkyl peripheral substitutions. While six-fold aromatic rings are predominant in the analysed structures, they also contain a significant fraction of five-fold rings, rare seven-fold rings and apparently no three- or four-fold rings. No two identically-structured asphaltenes were observed in the experiments and the structures are often non-planar. Among the analysed species organic radicals (CA12), charged species (CA3) and even an 18 ring "nano-graphene" (CA6) were observed\cite{JACS_2015_Asphaltenes} (see Table~\ref{table_asphaltenes}).  

Typically, asphaltenes exhibit a central aromatic core with peripheral alkane chains, in some cases the aromatic cores comprise several distinct polycyclic aromatic hydrocarbons (PAHs) connected by single bonds. Nevertheless, the most common type of structure is a single aromatic core with often long alkane side groups. 
Among the analysed species \cite{JACS_2015_Asphaltenes} there is no sign of any cage-like structures, probably because they contain too few C atoms, {\it i.e.}, $N_{\rm C} < 60$, and are therefore significantly smaller than fullerenes. 
It has been noted that molecular re-arrangements often occurred during the AFM measurements making an attribution of the structure of the non-planar side groups particularly  difficult, nevertheless methyl ($-$CH$_3$) appears to be the most abundant side group.\cite{JACS_2015_Asphaltenes} 
Further, in these experiments it was not possible to unambiguously determine the nature of the hetero-atoms/moieties in the P rings, which could be CH, CH$_2$, CO, N, NH or S. 

Using as a reference the superbly-detailed analytical work on the structure of over 100 asphaltene species,\cite{JACS_2015_Asphaltenes} and as summarised in Table~\ref{table_asphaltenes} using the above nomenclature (Section \ref{sect_nomenclature} above), we can derive some general observations about the types of aromatic-rich sub-structures likely to be found in asphaltenes: 
\begin{itemize}
\item Relatively compact (peri-condensed) structures dominate. 
\item Very extended (cata-condensed) structures are not common. 
\item Some show PAH islands, S$_{n=1-3}$, linked by single bonds. \\[-0.2cm]

\item PAs show more ring substitutions and P rings than CAs.
\item PAs exhibit longer side groups than CAs. 
\item PA side groups can be up to $\sim 2$\,nm long ($\lesssim 15$ C$-$C bonds). \\[-0.2cm] 

\item Central core S$_n$ domains can be up to $\sim 20$ rings,  {\it i.e.}, $\lesssim$ S$_{20}$.
\item For most S$_n$ domains $n$ is typically $\simeq 4-10$ rings.
\item S$_n$ domains show no methylene ($-$CH$_2-$) bridges. \\[-0.2cm]

\item S$_n$ domains are about as common as mixed S$_n$P$_q$(G) species. \\[-0.2cm]
 
\item P rings are generally single and bridging, {\it e.g.}, S$_n$PS$_m$
\item P rings can have methylene ($-$CH$_2-$) bridges, {\it i.e.}, P$^\prime$.
\item Most P rings are peripheral hetero-cycles, {\it i.e.}, P$^{\rm X}$.
\item P hetero-cycles are more abundant than S hetero-cycles.
\item P$^{\rm X}$ systems can be paired, {\it i.e.}, P$^{\rm X}$P$^{\rm X}$.
\item Double hetero-atom cycles are seen, {\it i.e.}, P$^{\rm XX}$.
\end{itemize}
In the experiments it was noted that the conformation of the analysed asphaltenes was affected by their adsorption on the surface, which tends to force them into a more planar configuration than the results would indicate.\citep{JACS_2015_Asphaltenes} 
Also, six-fold ring aromatic-dominated structures, S$_n$, are probably more easily analysed in the data and therefore may be somewhat over-represented in this large but still somewhat limited asphaltene sampling. 
Among the illustrated structures there does not appear to be a preference for any particular size of the S$_n$ domains, {\it i.e.}, all values of $n$ from 1 to 9 appear equally probable. 

In Table~\ref{table_asphaltenes} the characteristic structures of the analysed CA asphaltenes\citep{JACS_2015_Asphaltenes} are described using the nomenclature scheme proposed above (Section~\ref{sect_nomenclature}). From this table the proponderance of the S-rich aromatic moieties is clear, as is the bridging role of the P rings and their heterocyclic nature.

\begin{table*}
\caption{Some typical asphaltene structures ordered by increasing increasing number of carbon atoms in ring structures, N$_{\rm C}$, the number of rings , N$_{\rm R}$, and the ring variety, S $\rightarrow$ SP $\rightarrow$ GSP. The lines separating species with more than 30 C atoms from those with fewer, indicates the approximate limit for the size of the aromatic-rich species in the diffuse ISM.\citep[{\it e.g.},][]{1994ApJ...420..307J}}
\begin{center}
\begin{tabular}{lcccccl}
\hline
                     &                                     &                           &                     &                     &                           &                     \\[-0.25cm]
                     &                                      &                            &   C atoms &    methyl    &  alkyl   &     \\

   Asphaltene &                              &                                &  per ring  &    side groups        &   side groups &                  \\
   specimen   &   N$_{\rm C}$   &     N$_{\rm R}$   &  N$_{\rm C}$/N$_{\rm R}$ &    ( $-$CH$_3$ )        &   ( $-$R ) & Descriptor    \\
                     &                         &                            &                     &                     &                           &                     \\[-0.3cm]
\hline  
                                 &                        &                            &                     &                     &                           &                     \\[-0.25cm]
\multicolumn{4}{l}{\underline{S systems (six-fold rings)}}  \\[0.1cm] 
         CA36    &      22        &       5       &       4.4       &           1         &         2          &    S$_{4}$$-$S                                                     \\[0.05cm]
         CA74    &      22        &       6       &       3.7        &          0         &         1          &    S$_{6}$                                                             \\[0.05cm]
         CA49    &      24        &       6       &       4.0        &          0         &          2         &    S$_{4}$S$^{\rm X}$S                                        \\[0.05cm]
         CA12    &      25        &       6       &       4.2        &          1         &         0          &    S$_{6}$                                                            \\[0.05cm]
         CA47    &      25        &       7       &       3.6        &          0         &          1         &    S$_{7}^{\rm X}$                                                 \\[0.05cm]
         CA44    &      25        &       7       &       3.6        &          1         &          0         &    S$_{7}$                                                            \\[0.05cm]
         CA22    &      25        &       7       &       3.6        &          0         &          0         &    S$_{7}^{\rm X}$                                                 \\[0.05cm]
         CA65    &      26        &       7       &       3.7        &          0         &          0         &    S$_{7}$                                                             \\[0.05cm]
         CA1      &      28        &       8       &       3.5        &          1         &          0         &    S$_{8}$                                                            \\[0.05cm]
         CA9      &      28        &       8       &       3.5        &          1         &          0         &    S$_{8}$                                                            \\[0.05cm]
         CA61    &      28        &       8       &       3.5        &          0         &          0         &    S$_{8}$                                                             \\[0.05cm]
\hline
                                &                         &                     &                     &                     &                           &                     \\[-0.25cm]
         CA14    &      30        &       9       &       3.3        &          1         &          0         &    S$_{9}$                                                            \\[0.05cm]
         CA34    &      32        &       7       &       4.6        &          0         &          1         &    S$_{4}$$-$S$_{3}$                                          \\[0.05cm]
         CA7      &      32        &       9       &       3.6        &          1         &          0         &    S$_{9}$                                                            \\[0.05cm]
         CA79    &      34        &      10      &       3.4        &          1         &          3         &    S$_{10}$                                                           \\[0.05cm]
         CA5      &      35        &      11      &       3.2        &          0         &          2         &    S$_{11}$                                                          \\[0.05cm]
         CA72    &      40        &      11      &       3.6        &          0         &          3         &    S$_{11}$                                                           \\[0.05cm]
         CA6      &      58        &      18      &       3.2        &          0         &          0         &    S$_{18}$                                                          \\[0.05cm]
\hline
\hline
                                &                         &                     &                     &                     &                           &                     \\[-0.25cm]
\multicolumn{4}{l}{\underline{SP systems (six- and five-fold rings)}}  \\[0.1cm] 
         CA21    &      18        &      5      &     3.6       &          0         &          0         &    S$_{4}$P$^{\rm X}$                                          \\[0.05cm]
         CA23    &      20        &      5     &       4.0      &          2         &          0         &    SP$^{\rm X}$P$^{\rm X}$S$-$S                        \\[0.05cm] 
         CA41    &      20        &      5     &      4.0       &          0         &          1         &    S$_{3}$P$^{\rm X}$S                                        \\[0.05cm]
         CA95    &      20        &      6      &     3.3       &          2         &          0         &    S$_{5}^{\rm X}$P                                                \\[0.05cm] 
\hline
                                &                         &                     &                     &                     &                           &                     \\[-0.25cm]
         CA54    &      30        &      8      &     3.8       &          1         &          2         &    S$_{6}$P$^{\rm X}$S                                        \\[0.05cm]
         PA3      &       30       &      9      &      3.3      &          1         &         2          &    S$_{3}$P$^{\rm X}$P$^{\rm X}$S$_{2}$P$^{\rm X}$S  \\[0.05cm]
         CA48    &      31        &      9      &     3.4      &          1          &         0          &    S$_{6}$S$^{\rm X}$S$^{\rm X}$S                      \\[0.05cm]
         CA2      &      34        &      9      &     3.8       &          1         &          0         &    S$_{4}$PS$_{4}$                                            \\[0.05cm]
         CA13    &      34        &      9      &      3.8      &          0         &          0         &    S$_{7}$PS                                                       \\[0.05cm]
         CA39    &      40        &      8      &     5.0       &          1         &          0         &    S$_{6}$P$^{\rm X}$S                                        \\[0.05cm]
         CA91    &      42        &     12     &     3.5       &          0         &         4          &    S$_{11}$P$^{\rm XX}$                                    \\[0.05cm]
         CA16    &      44        &     11     &      4.0      &          0         &         0          &    S$_{3}$P$^{\rm X}$S$_{3}$$-$S$_{4}$            \\[0.05cm]
         CA60    &      46        &     14     &     3.3        &          0         &         4          &    S$_{9}$PS$_{2}$P$^{\rm X}$S                         \\[0.05cm]
         CA15    &      54        &     15     &     3.6       &           0        &          0         &    S$_{3}$PS$_{8}$P$^{\rm X}$S$_{2}$              \\[0.05cm]
\hline
\hline
                                &                         &                     &                     &                     &                           &                     \\[-0.25cm]
\multicolumn{4}{l}{\underline{GSP systems (seven-, six- and five-fold rings)}}  \\[0.1cm]
         CA3      &      40        &      12     &     3.3      &           3        &           0        &    S$_{5}$P$^\prime$GPS$_{2}$P$^\prime$S   \\[0.05cm]
\hline
                                &                   &                     &                     &                     &                           &                     \\[-0.25cm]
\end{tabular}  
\begin{list}{}{}
Notes:
\item[] 1. The number of C atoms in rings ignores pendant alkyl side groups ($-$R) such as $-$CH$_3$. 
\item[] 2. P$^\prime$ indicate five-fold (P) ring sites with single methylene, $-$CH$_2-$ substitutions.
\item[] 3. S$^{\rm X}$ and P$^{\rm X}$ (P$^{\rm X}$$^{\rm X}$) indicate peripheral six-fold (S) and five-fold (P) ring sites with single (double) hetero-atom substitutions ({\it i.e.}, hetero-cycles).
\item[] 4. The asphaltene PA3 is from virgin crude oil and did not undergo further treatment, all the other analysed asphaltenes (CAs) are derived from coal. 
\end{list}    
\end{center}    
\label{table_asphaltenes}
\end{table*}

Using the ideas gleaned from the (sub-)structures observed in the naturally-occurring aromatic-rich asphaltenes (Table~\ref{table_asphaltenes}) we propose that similar sub-structures are also likely to be present in [X-doped] interstellar hydrogenated amorphous carbon, a-C(:H[:X]), grains and especially in aromatic-rich a-C nanoparticles. 
We now explore the likely consequences of the presence of such sub-structures in nano-particles in the ISM.

%------------------------------------------------------------------
\section{Dehydrogenated and ionised aromatic structures} 
\label{sect_dehyd}
%------------------------------------------------------------------ 

It has long been proposed in astrophysical studies that interstellar aromatic species will be dehydrogenated and ionised under the harsh UV irradiation conditions of the diffuse ISM. However, it has been found that ionised and partially dehydrogenated aromatic species are less resistant to photo-destruction than the parent neutral molecules. For example, the ovalene cation, C$_{32}$H$_{14}^+$, has a photo-dissociation rate that is four times that of the neutral molecule and for the completely dehydrogenated form, C$_{32}^+$, it is three orders of magnitude larger.\cite{1996A&A...305..616A}  From this work it appears that only "PAHs" with $n_{\rm C} < 50$ would be significantly dehydrogenated. However, these same "PAHs" will be destroyed by C$_{2}$H$_{2}$ loss on a time-scale of years in regions of the ISM with high UV radiation field strengths (reflection nebulae, planetary nebu\ae, H{\footnotesize II} regions, \ldots) where the IR emission bands are observed.\citep{1996A&A...305..602A} 

However, the consequences of the loss of hydrogen atoms from aromatic carbonaceous species and the structure of the resulting radical ion have rarely been considered. It generally seems to be assumed that aromatic species, such as PAHs, retain their aromatic network structures and just lose hydrogen atoms from the periphery, without any structural re-arrangement. Further, it is assumed that hydrogen atom addition to the radical ion re-generates the original aromatic structure However, in the case of severely dehydrogenated PAHs, this view is probably rather simplistic and seems to be contradicted by experimental evidence.\citep[{\it e.g.},][]{1999IJMSp.185....1G} 

The stepwise dehydrogenation of a suite of PAHs\footnote{Fluorene, C$_{13}$H$_{10}$, phenanthrene, C$_{14}$H$_{12}$, pyrene, C$_{16}$H$_{10}$,  perylene, C$_{20}$H$_{12}$, benz[a]anthracene, C$_{18}$H$_{12}$, chrysene, C$_{18}$H$_{12}$, triphenylene, C$_{18}$H$_{12}$ and coronene, C$_{24}$H$_{12}$} to form C$_n$H$_x^+$ cations has been extensively studied using sustained off-resonance irradiation.\cite{1999IJMSp.185....1G} In these experiments the irradiation resulted in the sequential loss of single hydrogen atoms from the parent ions and the eventual formation of completely dehydrogenated carbon cluster ions, C$_n^+$, in the case of coronene, perylene, and benz[a]anthracene. 
En route to the formation of C$_{24}^+$, from C$_{24}$H$_6^+$, no intermediate C$_n$H$_x^+$ cations with $n = 1$, 2 or 4 were formed, and about two thirds of the C$_{24}$H$_3^+$ cations decompose through carbon atom and hydrocarbon fragment loss.
For smaller PAHs no carbon cluster ions were observed, most likely because of fragmentation. 
In the course of the experiments it was noted that C$_n$H$_5^+$ ions could be formed after the loss of five hydrogen atoms from pyrene and fluorene or seven hydrogen atoms from coronene, perylene, benz[a]anthracene, chrysene and triphenylene. 
The loss of further hydrogen atoms to form C$_{n}$H$_4^+$ or C$_n$H$_3^+$ ions  was only seen for benz[a]anthracene, chrysene, and triphenylene. 
Different dissociation pathways to C$_n$H$_x^+$ ions with odd and even numbers of hydrogen atoms were noted with even electron species C$_n$H$_{2m+1}^+$ dissociating more easily, via single hydrogen atom loss, to form more stable C$_n$H$_{2m}^+$ ions. 
The difficulty in forming C$_n$H$_x^+$ ($0 \leq x \leq 4$) by stepwise hydrogen atom elimination is assumed to be due to significant differences in the structure of cation radicals with less than five hydrogen atoms.\citep{1999IJMSp.185....1G}

%------------------------------------------------------------------
\subsection{Dehydrogenation pathways for aromatic moieties} 
\label{sect_dehyd_struct}
%------------------------------------------------------------------ 

% FIGURE 1 *********************************************************
\begin{figure}[!h]
 %\centering\includegraphics[width=5in]{Z_figures/PAH_dehydrog0.pdf}
 \centering\includegraphics[width=5in]{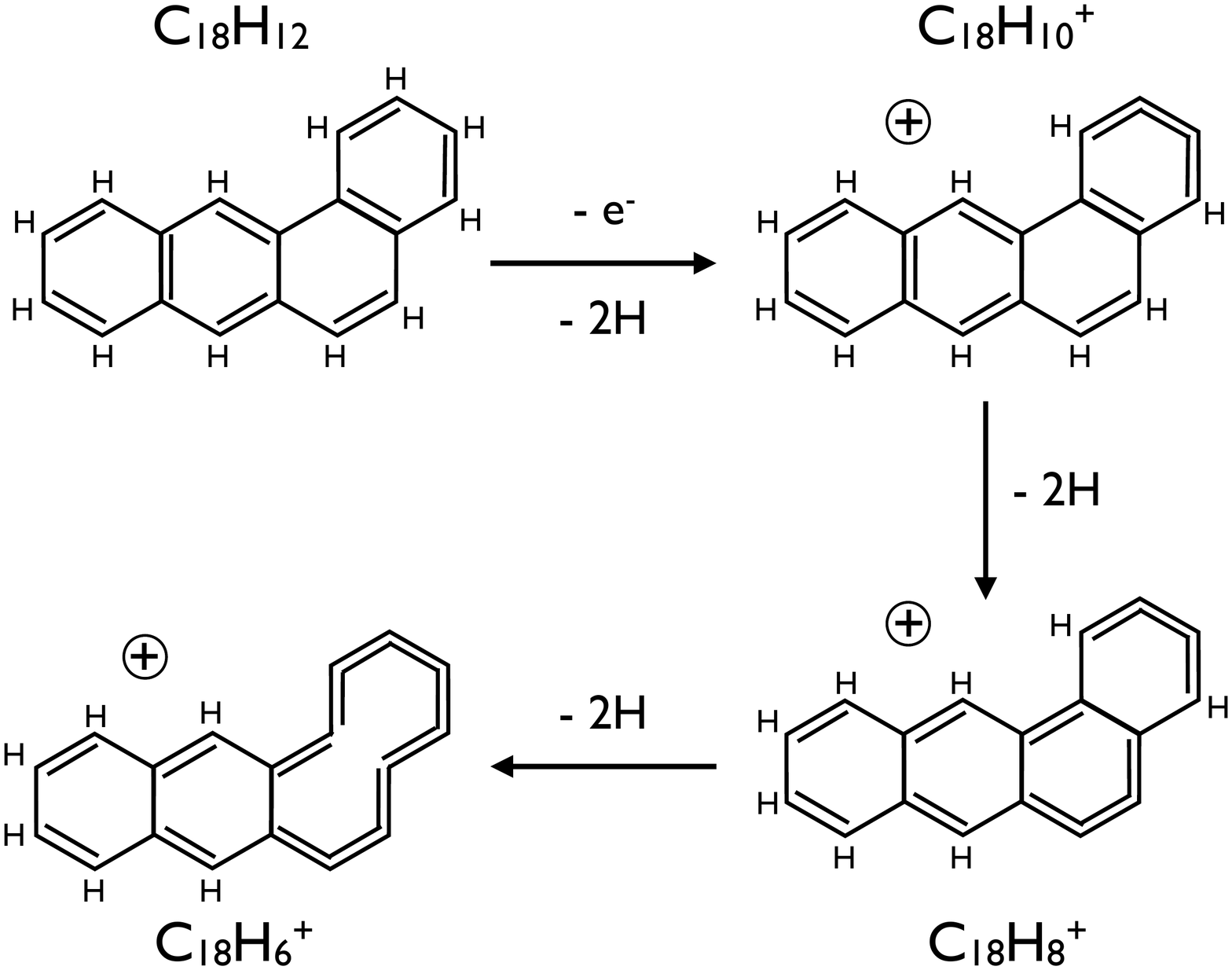}
 \caption{A proposed pathway for the dehydrogenation of C$_{18}$H$_{12}$ (S$_4$) benz[a]anthracene to C$_{24}$H$_6^+$.  A further single hydrogen atom dehydrogenation would then form a C$_{24}$H$_5^+$ ion.}
 \label{fig_PAH_dehydrog}
\end{figure}
% *********************************************************

% FIGURE 2 *********************************************************
\begin{figure}[!h]
 %\centering\includegraphics[width=5in]{Z_figures/PAH_dehydrog1.pdf}
 \centering\includegraphics[width=5in]{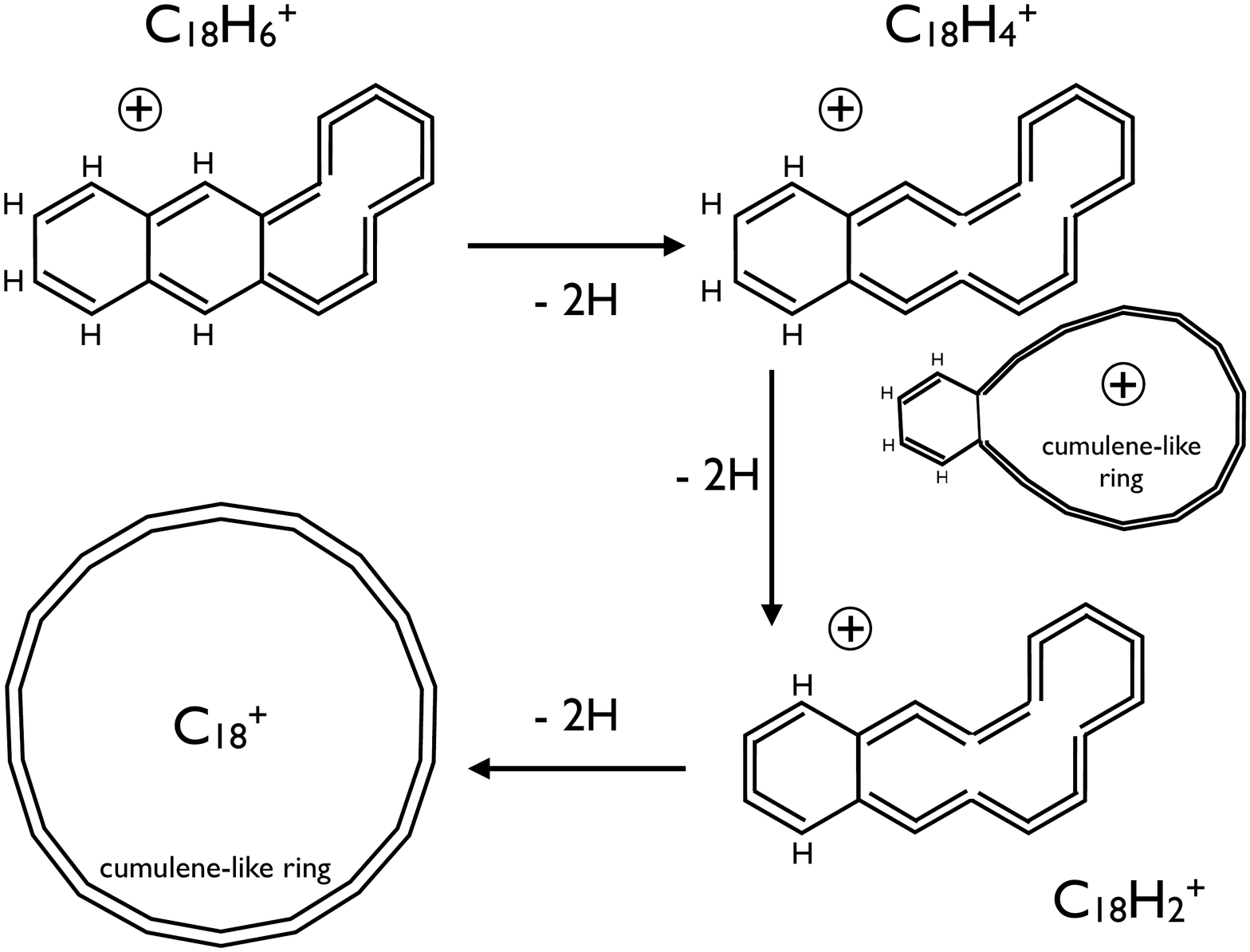}
 \caption{A proposed pathway for the  dehydrogenation of the C$_{18}$H$_{6}^+$ cation to C$_{24}^+$. The double ring structure in the upper and middle right shows the most likely bicyclic form of the C$_{18}$H$_{4}^+$ cation, {\it i.e.}, an ortho-substituted six-fold ring with a 12 carbon atom  cumulene-type ring \citep{double_ring_dehydPAH2,double_ring_dehydPAH1,1999IJMSp.185....1G}.}
 \label{fig_PAH_dehydrog2}
\end{figure}
% *********************************************************

The dehydrogenation of molecular PAH ions most likely proceeds by the elimination of two immediately adjacent (ortho) hydrogen atoms leading to a stable aryne bond. The aryne bond was long-assumed to be acetylenic in nature.\citep[{\it e.g.},][]{acetylenic_aryne_1992} However, more recent work, highlighted in an article entitled ``First snapshot of elusive intermediate'' (Philip Ball, Chemistry World, September 2015, p. 26) highlights the results of this work.\cite{Gross_etal_2009,Schuler_etal_2014,2015NatCh...7..623P} Using atomic resolution imaging these researchers have shown that the aryne bond is actually a non-linear C=C=C=C "cumulene-like" structure.\footnote{A perhaps somewhat analogous structure in a-C(:H) nano-particles could be the sites for the pair-wise elimination of adjacent hydrogen atoms, from the aliphatic/olefinic bridges linking the intrinsic aromatic domains, as recently suggested as a viable route to H$_2$ formation in moderately-excited PDRs.\citep{2015A&A...581A..92J}} 

For the C$_{18}$H$_{12}$ PAHs benz[a]anthracene, chrysene and triphenylene dehydrogenation proceeds by a similar route and leads to the bicyclic product, C$_{18}$H$_4^+$, which is apparently more stable than linear and single cycle species.\citep{double_ring_dehydPAH1} The most stable bicyclic product is thought to contain a benzene ring ortho-substituted by a carbon loop.\citep{double_ring_dehydPAH2,double_ring_dehydPAH1} 
Fig.~\ref{fig_PAH_dehydrog} shows a likely dehydrogenation sequence for the C$_{18}$H$_{12}$ PAH benz[a]anthracene to the C$_{18}$H$_6^+$ ion, which proceeds via the progressive dehydrogenation of the three-ring phenanthrene sub-structure to form the cumulene-like structures noted above. Chrysene and triphenylene also contain the phenanthrene sub-structures and so the progressive dehydrogenation of these molecules to C$_{18}$H$_6^+$ ions will most likely follow the same sequence to the same product ion. In each of these cases the C$_{18}$H$_6^+$ ion has been dehydrogenated to a three-ring structure through cumulene-like chain formation, {\it i.e.}, the formation of double bonded C=C=C=C chains and eventually a 10 carbon atom cumulene-like chain (lower left structure in Fig.~\ref{fig_PAH_dehydrog}). The problem with this scenario is that it does not easily explain why the C$_{18}$H$_6^+$ ion structure is somewhat special because six-fold ring break-up has already started. The further dehydrogenation of this form of the cation could in any event proceed along the pathways shown in Fig.~\ref{fig_PAH_dehydrog2} to yield  the product bicyclic C$_{18}$H$_4^+$ ion \citep{double_ring_dehydPAH2,double_ring_dehydPAH1} and to a possible structure for the C$_{18}^+$ carbon cluster. Indeed, this type of radicalised double aromatic ring opening transformation, well-known as Bergman cyclisation, was recently revealed in its full structural glory through a beautiful scanning tunnelling microscopy experiment.\cite{Schuler_etal_2016}

% FIGURE 3 *********************************************************
\begin{figure}[!h]
 %\centering\includegraphics[width=5in]{Z_figures/PAH_dehydrog2.pdf}
 \centering\includegraphics[width=5in]{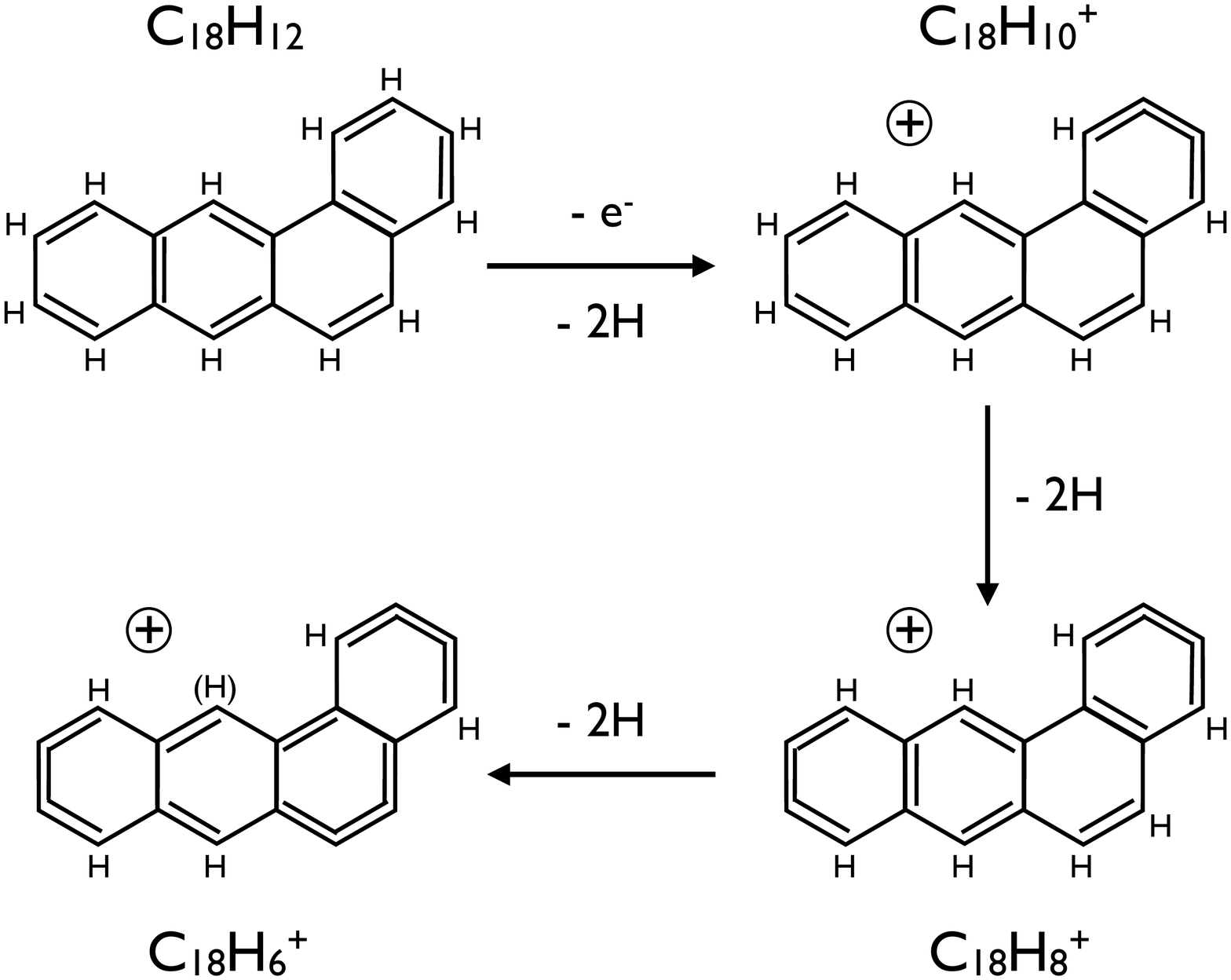}
 \caption{An alternative pathway for the dehydrogenation of C$_{18}$H$_{12}$ (S$_4$) benz[a]anthracene to C$_{24}$H$_6^+$. The bracketed hydrogen atom, (H), in the lower right ion indicates a site than could be dehydrogenated to form a C$_{24}$H$_5^+$ ion.}
 \label{fig_PAH_dehydrog3}
\end{figure}
% *********************************************************

% FIGURE 4 *********************************************************
\begin{figure}[!h]
 %\centering\includegraphics[width=5in]{PAH_dehydrog3.pdf}
 \centering\includegraphics[width=5in]{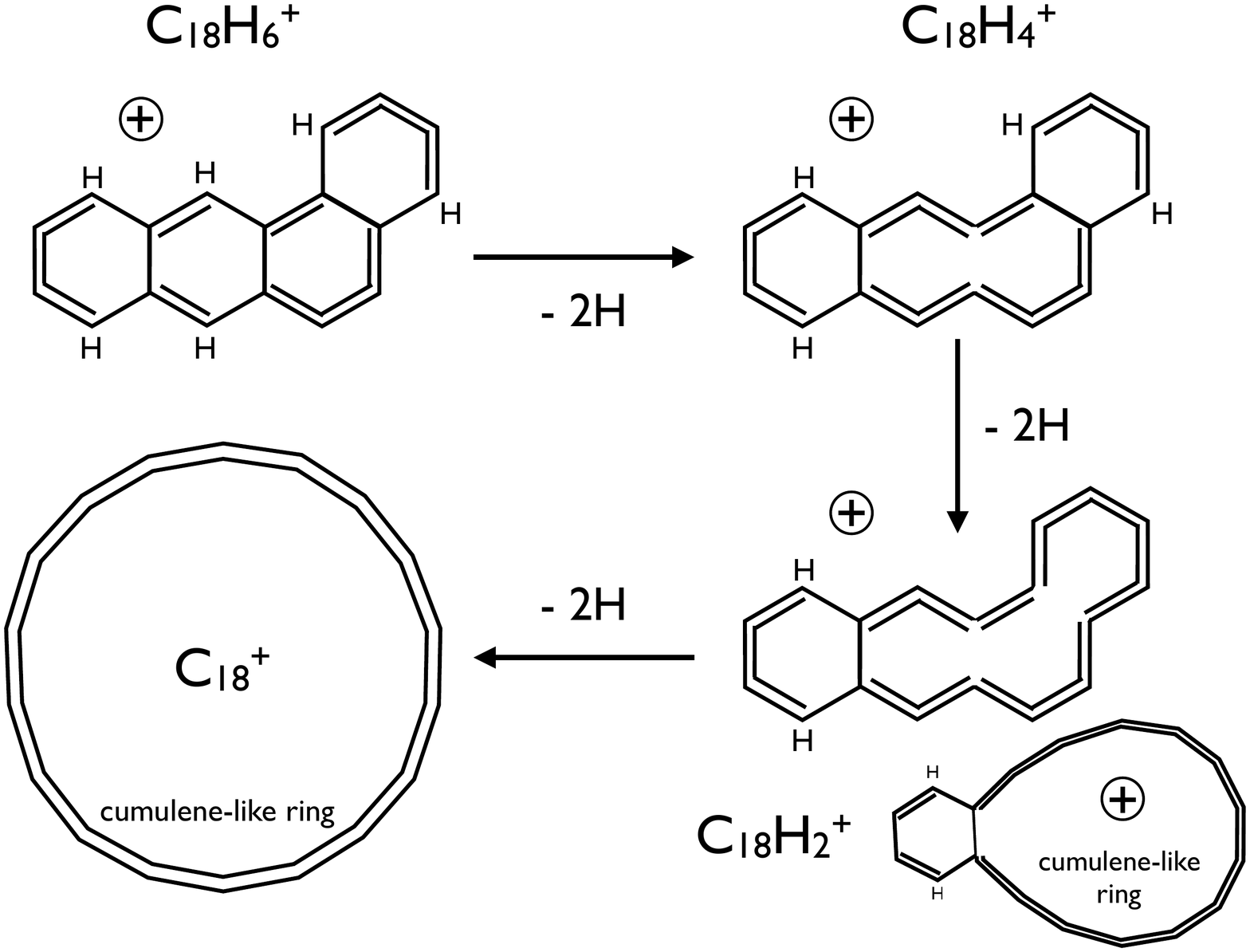}
 \caption{The follow-on to the alternative dehydrogenation pathway for theC$_{18}$H$_{6}^+$ cation to C$_{24}^+$. In this case the formation of a differernt bicyclic cation, C$_{18}$H$_{2}^+$, occurs somewhat later in the dehydrogenation sequence.}
 \label{fig_PAH_dehydrog4}
\end{figure}
% *********************************************************

It appears that an alternative pathway involving the preferred and sequential formation of C=C=C=C cumulene-like structures first may also provide a viable route to the C$_{18}$H$_6^+$ cation  (see Fig.~\ref{fig_PAH_dehydrog3}) without the need to break six-fold rings. In this case the dehydrogenation has seemingly gone about as far as it can go along this path without a major disruption of the original PAH structure. In this case a critical-state ion structure has been formed, for which the further dehydrogenation of this C$_{18}$H$_6^+$ cation (lower left structure in Fig.~\ref{fig_PAH_dehydrog3}) can only proceed via the formation of 
cumulene-like chains through the disruption of six-fold ring structures. 
Fig.~\ref{fig_PAH_dehydrog4} shows a likely dehydrogenation sequence for a cumulene-dominated form of the benz[a]anthracene-derived ion C$_{18}$H$_6^+$ to an alternative product bicyclic structure, C$_{18}$H$_2^+$, and to the same form of the C$_{18}^+$ carbon cluster as for the previously-described pathway shown in Figs.~\ref{fig_PAH_dehydrog} and \ref{fig_PAH_dehydrog2}. 
Using simple bond energy considerations, {\it i.e.}, summing the energy due to the breaking of C$-$H bonds and the conversion of C$-$C bonds into C=C bonds, it can be shown that the alternative pathway to the formation of the C$_{24}$H$_{6}^+$ cation (Fig.~\ref{fig_PAH_dehydrog3}) appears to be more favourable by almost 4\,eV.

Regardless of the pathway that PAH dehydrogenation actually follows Figs.~\ref{fig_PAH_dehydrog} to \ref{fig_PAH_dehydrog4} show that significant changes in structures are to be expected. In particular, the pathway shown in Figs.~\ref{fig_PAH_dehydrog3} and \ref{fig_PAH_dehydrog4} shows a major structural difference between the cations with 5 or 6 hydrogen atoms (Fig.~\ref{fig_PAH_dehydrog3}) and those with less than 5 hydrogen atoms (Fig.\ref{fig_PAH_dehydrog4}), which could perhaps better explain the difference in the dehydrogenation behaviour of these cations. However, in this case the products formed prior to complete dehydrogenation are a tricyclic C$_{18}$H$_4^+$ ion or a bicyclic C$_{18}$H$_2^+$ ion rather than a bicyclic C$_{18}$H$_4^+$ cation.\cite{double_ring_dehydPAH2, double_ring_dehydPAH1,1999IJMSp.185....1G}  

Nevertheless, and whatever the exact dehydrogenation route, it seems that in all cases the dehydrogenation of PAH-like structures proceeds by single hydrogen atom elimination down to C$_n$H$_5^+$ but thereafter involves significant carbon atom loss from the structure, {\it i.e.}, by a fragmentation of the parental PAH framework. In some cases (C$_{n}$H$_x^+$; $n \geq 20$, $x \leq 10$) the cation species do perhaps retain a memory of the parent PAH, suggesting that some PAH structures may be only slightly modified by low energy hydrogen elimination.\citep{1999IJMSp.185....1G} Thus, it appears that the exact routes for PAH dehydrogenation and cation formation are particularly dependent upon the details of the aromatic structure and perhaps the symmetry of the parent PAH, with highly symmetric PAHs following a different path from their asymmetric and cata-condensed sisters. 

It has been found that adding hydrogen to C$_n^+$ carbon cluster cations, with $n = 9$ to 22, leads to a change in the structure from monocyclic C$_n^+$ to linear C$_{n}$H$_x^+$ forms ($x \leq 3$).\cite{double_ring_dehydPAH2,double_ring_dehydPAH1}  In addition a new but minor bicyclic isomer appears from $x = 4$ and $n \geq 15$, which increased in abundance with $n$ at the expense of the monocyclic and linear forms. Further, the bicyclic forms of the C$_{16}$H$_4^+$ ion appear to be much more stable than either the monocyclic or linear forms.\cite{double_ring_dehydPAH1} 

\begin{figure}[!h]
 %\centering\includegraphics[width=5in]{Z_figures/coronene_dehydrog0a.pdf}
 \centering\includegraphics[width=5in]{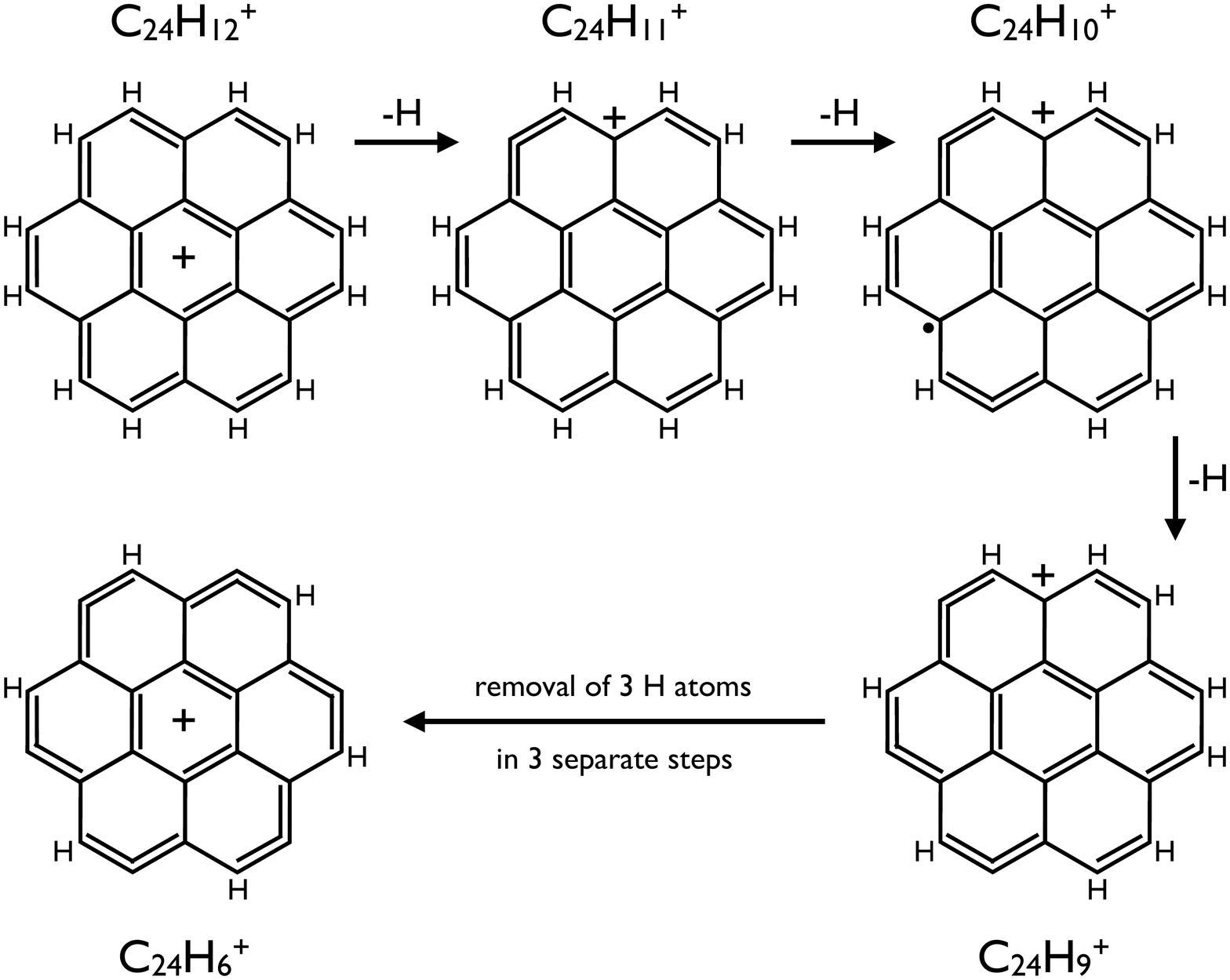}
 \caption{One possible pathway for the  dehydrogenation of the C$_{24}$H$_{12}^+$ (S$_7$) coronene cation to C$_{24}$H$_{6}^+$.}
 \label{fig_coronene_dehydrog1}
\end{figure}

\begin{figure}[!h]
 %\centering\includegraphics[width=5in]{Z_figures/coronene_dehydrog0b.pdf}
 \centering\includegraphics[width=5in]{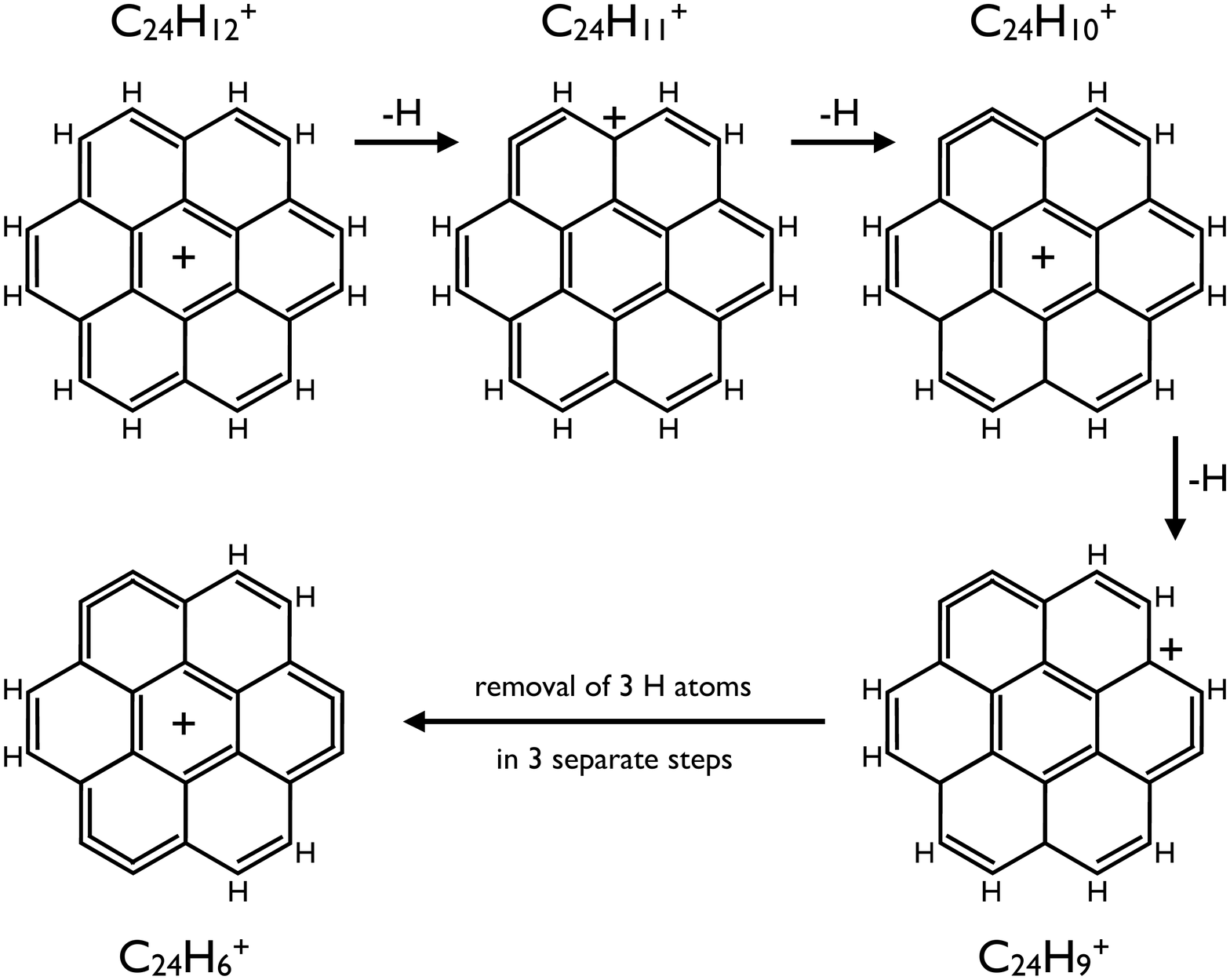}
 \caption{A second possible pathway for the  dehydrogenation of the C$_{24}$H$_{12}^+$ (S$_7$) coronene cation leading to C$_{24}$H$_{6}^+$ structure with a different but equally symmetric structure.}
 \label{fig_coronene_dehydrog2}
\end{figure}

\begin{figure}[!h]
 %\centering\includegraphics[width=5in]{Z_figures/coronene_dehydrog1.pdf}
 \centering\includegraphics[width=5in]{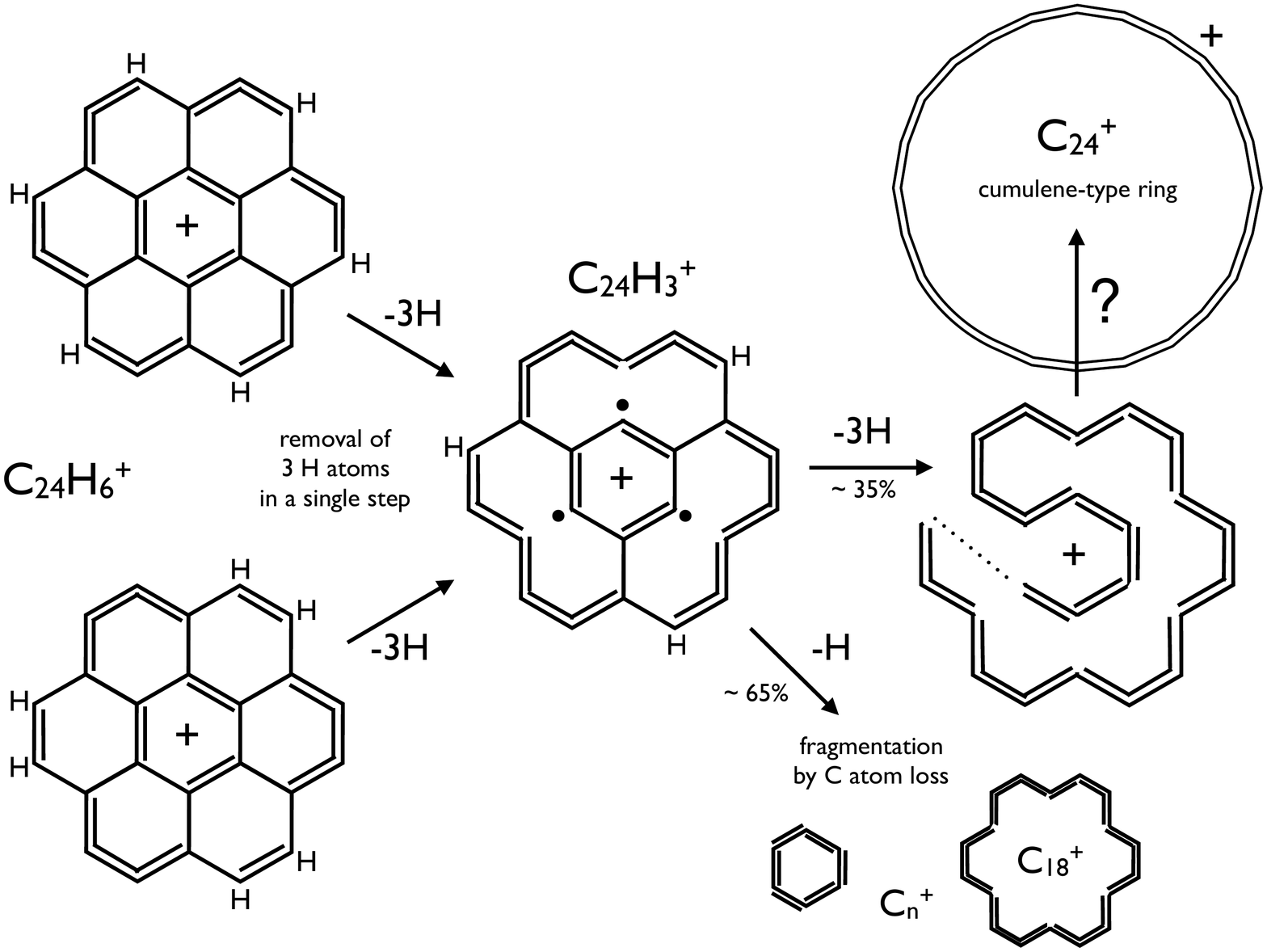}
 \caption{A possible pathway for the  dehydrogenation of the C$_{24}$H$_{6}^+$ ion leading to C$_{24}^+$, which is assumed to have a cumulene-type ring structure.}
 \label{fig_coronene_dehydrog3}
\end{figure}

The dehydrogenation of the larger coronene molecule must also follow a similar sequence but with slight differences indicated by the inability to form certain configuratrions along the way.\citep{1999IJMSp.185....1G}  
Figs.~\ref{fig_coronene_dehydrog1} and \ref{fig_coronene_dehydrog2} show possible pathways for the single H atom stepwise dehydrogenation of the coronene cation C$_{24}$H$_{12}^+$ to C$_{24}$H$_{6}^+$. The two pathways lead to equally symmetric cations but to quite different structures. In both of the suggested pathways, and up to this point, the seven ring structure of the parent coronene molecule is preserved. 
The further dehydrogenation of the C$_{24}$H$_{6}^+$ cation can only proceed by the breaking of one of the six-fold rings. As mentioned above the formation of the carbon cluster C$_{24}^+$, by the dehydrogenation of the C$_{24}$H$_6^+$ ion, only forms an intermediate C$_{24}$H$_3^+$ cation  and a large fraction ($\sim \frac{2}{3}$) of these ions actually fragment en route to C$_{24}^+$.\cite{1999IJMSp.185....1G}  Thus, the intermediate species with $x = 1$, 2 or 4 are either highly unstable or are bypassed by the dehydrogenation-imposed re-structuring, {\it i.e.}, the simultaneous breaking of three six-fold ring bonds to form C$_{24}$H$_3^+$ and then a further 4 six-fold ring bonds. 
Fig.\ref{fig_coronene_dehydrog3} shows a likely scenario for the C$_{24}$H$_x^+$ ($x < 6$) cation re-structuring leading to C$_{24}$H$_3^+$ and ultimately C$_{24}^+$ during dehydrogenation. 

% FIGURE 5 *********************************************************
% *********************************************************

% FIGURE 6 *********************************************************
% *********************************************************

% FIGURE 7 *********************************************************
% *********************************************************

%------------------------------------------------------------------
\subsection{Some consequences of dehydrogenation} 
\label{sect_dehyd_consequences}
%------------------------------------------------------------------ 

The re-structuring of aromatic moieties during dehydrogenation, via cumulene-like chain and ring formation, could have some interesting consequences for the UV photo-processing of the sub-structures in a-C(:H) nano-particles in photo-dissociation regions (PDRs) and H{\footnotesize II} regions. Clearly the aromatic sub-structures intrinsic to the grains can be dehydrogenated, as described above, but following a recently proposed H$_2$ formation mechanism it is most likely that the aliphatic/olefinic bridges between the aromatic moieties are UV processed before the aromatic domains are destroyed.\cite{2015A&A...581A..92J} This work also suggests that the UV processing of the bridges can take them as far as olefinic species but it now seems that their complete dehydrogenation could result in cumulene-like structures that will most likely be formed just before or in the process of nano-particle fragmentation. 

Hydrogen atom addition to a cumulene-type ring is known to lead to linear species, {\it i.e.}, ring-breaking.\citep{double_ring_dehydPAH2,double_ring_dehydPAH1} Thus, it is likely that the severe dehydrogenation of aromatic species in the ISM leads to structures that are very different from the parent structure. The reaction of gas phase hydrogen atoms with these dehydrogenated cyclic or bridging species likely does not necessarily lead to the re-formation of the original structure, through re-hydrogenation, but rather to linear molecule formation and/or the fragmentation of the chains. 

In the following some of the consequences of UV processing in intense radiation field regions leading to cumulene-type structure formation are considered. The following schematic indicates how cumulene-like bridging species, some with nitrogen hetero-atoms, could form and how they might react with atomic hydrogen (the "dangling" bonds illustrated as $-$, $=$ and $\equiv$ are assumed to be connected to the larger contiguous grain structure):
\[
-{\rm CH}_2-{\rm CH}_2-{\rm CH}_2-{\rm CH}_2- \ \ \ \ - \ 4{\rm H} \ \ \ \rightarrow \ \ \ \ \ \ \ \ \ \ \ \ \ \ \ \ \ \ \ \ \ \ \ \ \ 
\]
\[
-{\rm CH}={\rm CH}-{\rm CH}={\rm CH}- \ \ \ \ \ \ \ \ \ \ \ \ - \ 4{\rm H} \ \ \ \rightarrow \ \ \ \ \ \ \ \ \ \ \ \ \ \ \ \ \ \ \ \ \ \ \ \ \
\]
\[
={\rm C}={\rm C}={\rm C}={\rm C}=  \ \ \ \ \ \ \, + \ \ \ {\rm H} \ \ \ \rightarrow \ \ \ {\rm H}-{\rm C}\equiv{\rm C}-{\rm C}\equiv{\rm C}- \ \  
\]
\begin{equation} 
={\rm C}={\rm C}={\rm N}-{\rm CH}= \ \ \ + \ \ {\rm H} \ \ \ \rightarrow \ \ \ \equiv{\rm C}-{\rm C}\equiv{\rm N} \ + \ {\rm H_2C}=
\end{equation}
The above sub-structures could be parts of cumulene-type rings or the dehydrogenated bridging links between the aromatic domains in a-C(:H) nano-particles. Perhaps these species and their reaction with atomic hydrogen could provide an alternative top-down route to (cyano)polyyne molecule formation in the ISM?

%------------------------------------------------------------------
\section{A new form of carbon molecule?} 
\label{sect_new_form}
%------------------------------------------------------------------ 

% FIGURE 8 *********************************************************
\begin{figure}[!h]
 %\centering\includegraphics[width=5in]{Z_figures/couroneanes.pdf}
 \centering\includegraphics[width=5in]{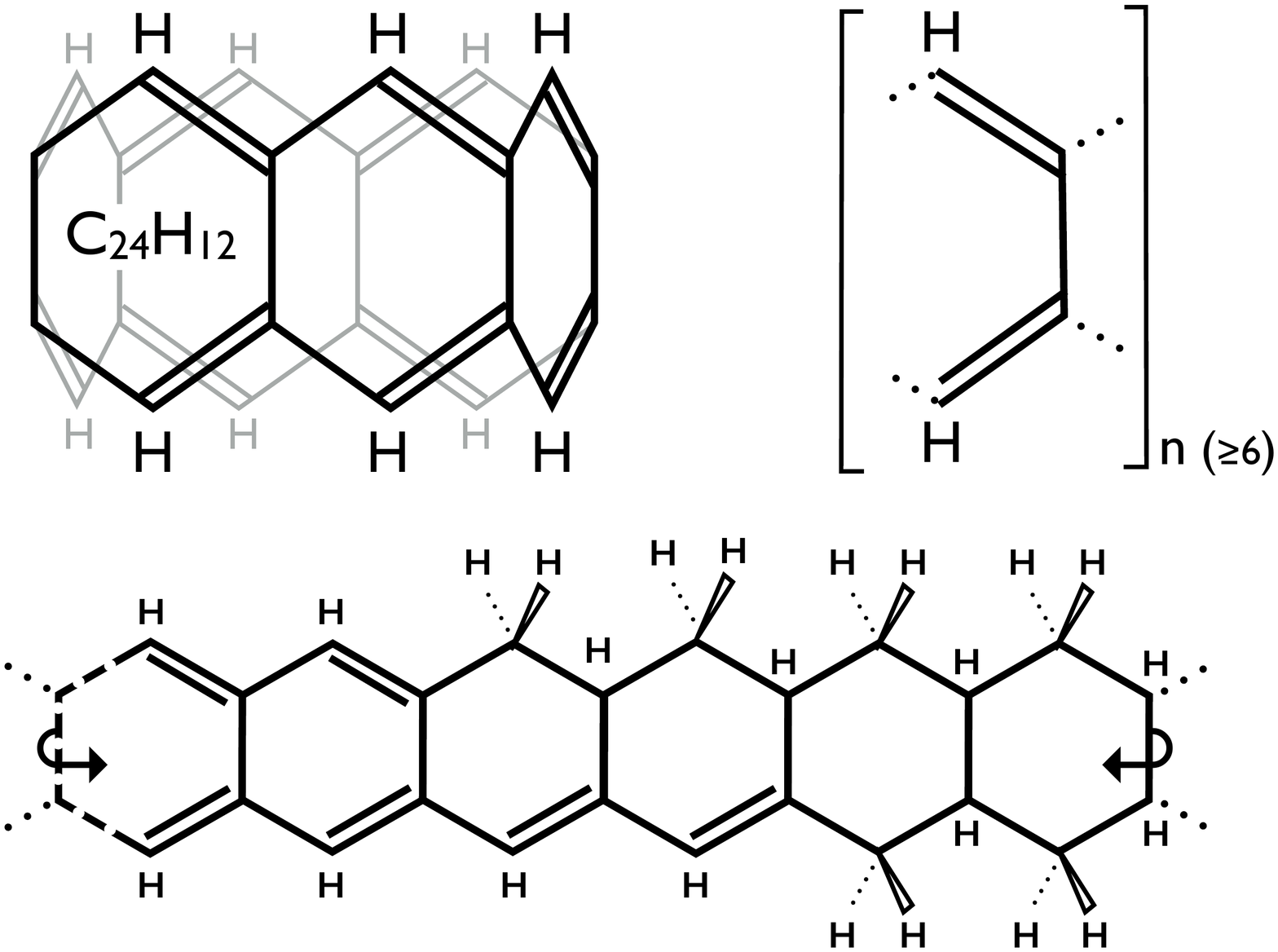}
 \caption{A proposed ``ring of rings" family of molecules: couronene, C$_{24}$H$_{12}$ (S$_6$, upper left), the structure for a concatenated cycle of aromatic rings with generic formula (C$_{4}$H$_{2})_n$ (S$_n$, upper right). If all the carbon atoms are $sp^3$ then the aliphatic multi-ring structure is a couronane, C$_{24}$H$_{36}$, generically (C$_{4}$H$_{6})_n$, where $n \geq 6$. Mixed aliphatic/aromatic ring structures would then be courone-anes, with the generic formula (C$_{4}$H$_{2-6}$)$_n$, where $n \geq 6$.}
 \label{fig_couronene}
\end{figure}
% *********************************************************

A coronene isomer and a new family of aromatic structures (C$_4$H$_2$)$_{n}$, with $n \geq 6$ is proposed; the name couronenes  is suggested for these molecules because of their crown-like structure. These structures are in fact carbon nano-tubes with a cylinder length of only one six-fold ring. However, they form a closed ring rather than the spiral structure often found in carbon nano-tube structures. Couronenes could be particularly interesting because the upper and lower crown edges are always stable annulene structures, which can likely be hydrogenated to form aliphatic couronanes (C$_4$H$_4$)$_{n}$, with $n \geq 6$, or mixed couronenes/couronanes (C$_4$H$_3$)$_{n}$, with $n \geq 6$ (courone-anes), with upper and lower crown edges having olefinic and aliphatic structures, respectively, or mixed olefinics and aliphatics on each edge. Fig.~\ref{fig_couronene} shows some examples of the likely structures of this proposed family of crown-like molecules. These species ought to have tuneable properties because, in principle, any number of rings can be joined in this way. Although, there must be some sterically-determined minimum ring number limit, which could mean that the minimum couronene/couronane molecules contain no less than 24 carbon atoms, {\it i.e.}, $n \geq 6$. Seemingly the addition of one or more nitrogen atoms into the structure, in a number of possible structural isomers, might to lead to some interesting transitions in the visible. 

Further, if the upper and lower edges of the crown can each be closed with a cap composed of a six-fold ring and six accompanying pentagons, then the result will be a closed C$_{36}$ fullerene structure. 

How might such couronene species be formed in the ISM, if indeed they are present there? The most likely route would be in the same inefficient way that fullerenes are formed, {\it i.e.}, via the UV photon interactions that lead to the vibrational excitation and the photo-processing dehydrogenation of a-C(:H) nano-particles.\citep{2012ApJ...761...35M} However, as for fullerenes, this route to their formation is likely to be equally inefficient because most of the possible parent species will by-pass these stable configurations and therefore only a small fraction of the carbon in dust in the ISM could be incorporated into couronenes, if they even exist there. 

However, independent of whether courone-ane species exist in the ISM their intrinsic properties are perhaps sufficiently interesting enough to warrant the further exploration of this idea.

%------------------------------------------------------------------
\section{The nature of the DIB carriers} 
\label{sect_ISM}
%------------------------------------------------------------------ 

Given that we have yet to find an explanation for more than $99.5$\% of the DIBs it would seem that some theoretical considerations, guided by experimental results, might yet be able to bring something to bear on the DIB problem by providing ideas about sub-sets of likely species that could then be explored in the laboratory. 
With this aim in mind this section aims to  provide a framework to elucidate the likely fundamental nature of the DIB carriers.

%------------------------------------------------------------------
\subsection{The observed properties of the DIB carriers} 
\label{sect_DIBobs}
%------------------------------------------------------------------ 

Here a brief and cursory summary of the major observational constraints on the DIBs is given. The interested reader should refer to the most recent proceedings and the more recent literature on the subject for a more complete overview of the subject.\citep[{\it e.g.},][]{1995ARA&A..33...19H,2014IAUS..297} 

There are over 400 known and well-determined DIBs and of these $\sim 100-200$ are common bands, {\it i.e.}, they are observed along every line of sight where the DIBs are seen, and they show a wide range of widths and intensities. 
For the handful considered so far, strong DIBs are not polarised. There do appear to be families of bands that are similar from one line-of-sight to another but these are very generic. 

The observational aspects of the DIBs presented here were drawn from an obviously limited selection of the enormous literature on the subject.\citep[{\it e.g.},][]{
1995ARA&A..33...19H,
2008A&A...484..381W,
2009A&A...498..785K,
2009ApJ...705...32H,
2010MNRAS.402.2548K,
2010ApJ...708.1628M,
2011ApJ...727...33F,
2011A&A...531A..68K,
2011ApJ...735..124K,
2013ApJ...773...42O,
2014ApJ...792..106W,
2015ApJ...800..137H,
2015MNRAS.452.3629L} 
In general, several key correlation and trends between the DIBs and environmental factors are seen, including that:  
\begin{itemize}
\item strong DIBs trace the diffuse atomic, HI, gas,  
\item they show a positive correlation with $N_{\rm HI}$,  
\item the show all possible correlation behaviours with $N_{\rm H_2}$,  
\item they are associated with the diffuse rather than the dense ISM, \\[-0.2cm]
\item they are broader than atomic or molecular lines,   
\item they are weaker in strong UV radiation fields,  
\item they show varied stability to UV radiation,   
\item broader DIBs are less sensitive to UV radiation,   
\item some NIR DIBs correlate and may be due to cations, 
\item inter-DIB correlations do not always go through zero, \\ {\it i.e.}, some DIBs are (compositionally) inter-dependant,   
\item they generally do not correlate well with each other, \\[-0.2cm]
\item they correlate with dust, {\it i.e.}, in extinction, $E(B-V)$, and emission, $12\,\mu$m continuum, which is itself a proxy for $N_{\rm H}$(total), 
\item they show a weak negative correlation with the 217\,nm UV bump, 
\item they show a weak negative correlation with the FUV extinction, 
\item weak DIBs correlate with enhanced FUV extinction, 
\item they show a "skin" effect, {\it i.e.}, strength decreasing above some $E(B-V)$ value, \\[-0.2cm]
\item they correlate with small molecules/radicals (but less well than with dust),  
\item some appear to be somehow associated with C$_2$, C$_3$, CN, CH, \ldots \\ 
but the same DIBs can be seen both with and without them,    
\item very few DIBS ($\sim 12$ DIBs, {\it i.e.}, $<  5$\%) show PQR branches, perhaps indicating prolate top molecules with $\lesssim 5-7$ atoms, 
\end{itemize}
These characteristics seem to strongly indicate that the DIB carriers are most likely formed by a top-down fragmentation process because a formation by bottom-up chemistry would lead to the same chemistry along all lines-of-sight. However, the same DIBs are observed along lines of sight  with and without the same molecules, radicals and ions. For example, in a study of 414 DIBs in the $\lambda = 390-810$\,nm region towards HD\,183143 it was found that the DIBs are redder, broader and weaker and that there is no C$_2$, whereas the line-of-sight towards HD\,204827 shows the same DIBs but high $N({\rm C}_2)/E(B-V)$.\cite{2009ApJ...705...32H}  
However, there appear to be weak features along both lines of sight and most of the DIBs are present towards both stars, albeit with very different relative strengths.\footnote{D. E. Welty, private communication} 

Interestingly, PAH cations are known to absorb in the visible and near-IR but no neutral PAHs or PAH cations have yet been detected in diffuse ISM and such species cannot therefore be the carriers of the DIBs.\citep[{\it e.g.},][]{2011MNRAS.412.1259G,2011A&A...530A..26G,2011ApJ...728..154S} This is consistent with work that shows that neutral and ionised PAHs cannot be the carriers of DIBs.\cite{2011ApJ...742....2S}  
Further, it has been shown that $l-$C$_3$H$_2$ is not a DIB carrier and it is therefore likely that this and other small linear molecules are not the origin of the DIBs.\cite{2011ApJ...735..124K}

%------------------------------------------------------------------
\subsection{Theoretical constraints on the DIB carriers} 
\label{sect_DIBconstraints}
%------------------------------------------------------------------ 

The DIBS are almost certainly due to the electronic transitions of the carrier species, which could be molecules, polyatomic radicals and/or ions. 
As has clearly been shown the 578.0\,nm DIB towards Hershel~26 is most likely due to a polar molecule with $n \lesssim 7$.\cite{2013ApJ...773...42O}  Thus, small $5-7$ heavy atom polar species are the most likely culprits for some DIBs, at least for those that show PQR-type  branch structures. Perhaps if the carrier species of the 578.0\,nm DIB are cyclic then they could contain about double this number of atoms ({\it i.e.}, $10-15$ heavy atoms). The conundrum here is that such small species are likely to have a rather short lifetime in the diffuse ISM, where only (aromatic-rich) species with more than about $30-50$ carbon atoms appear to be stable.\cite{1996A&A...305..616A} 

It was recently proposed that carbonaceous nano-particles could provide a viable explanation for the DIBs.\cite{2014P&SS..100...26J}  
In this case the hetero-atom doping of a-C(:H) (nano-)particles with O, N  $\gg$  Mg, Si  $>$  S  $\gg$  P, in order of cosmic abundance, or N, P, O, S  $>$  Mg, Si, in order of their chemical affinity for (aromatic) ring systems, perhaps provides a promising framework for further exploration. Although these particles would appear to have many advantages as viable DIB candidates, if they are too big they may actually be too stable to explain the sensitivity of some of the DIBs to UV radiation. 

The number of possible structural isomers for a given chemical composition C$_n$H$_m$X$_y$ is enormous; for species with $\gtrsim 30$ heavy atoms it is in the millions. The key (sub-)structures then probably contain only of the order of tens of carbon atoms because any more than this and the number of isomeric variations would be large and lead to millions of DIBs rather than the observed hundreds. 
Hence, the key (sub-)structures in the species that produce the DIBs must lie in a somewhat restricted set of chemical configurations. 

If the DIB carriers were to be formed by a top-down fragmentation process, say by the photolysis or photo-fragmentation of a-C(:H) nano-particles,  then they are most likely to be cyclic species because these are probably the most stable (sub-)structures that would be released from a-C(:H) nano-particles during UV photolysis. Further, a top-down process would provide an en route triage that weeds out the less stable structures, such as the olefinic/aliphatic links between the aromatic domains,\cite{2012A&A...542A..98J,2012A&A...540A...2J,2012A&A...540A...1J} which will in any event be rapidly photo-dissociated in the diffuse ISM \citep[{\it e.g.},][]{2012A&A...542A..98J,2013A&A...555A..39J,2014P&SS..100...26J} as discussed above.

%------------------------------------------------------------------
\subsection{The size and form of the DIB carriers} 
\label{sect_DIBstruct}
%------------------------------------------------------------------ 

In the diffuse ISM aromatic species with $\lesssim 30$ heavy atoms (C, N, O, \ldots) will likely rapidly be destroyed by UV photons.\citep[{\it e.g.},][]{1994ApJ...420..307J} However, when the dehydrogenation and ionisation states are taken into account\citep{1996A&A...305..616A} it appears that only species with $\gtrsim 50$ `heavy' atoms will be resistant to destruction in the regions of the ISM where the IR emission bands are observed.\citep{1996A&A...305..602A} 
This seems to be consistent with the observed structure in some DIBs being consistent with the rotational contours of rather small ($5-7$ heavy atom)\cite{2013ApJ...773...42O} gas phase molecules.\footnote{It was previously reported for the same set of DIBs that the carriers must be $20-40$ heavy atom gas phase molecules.\citep{1976MNRAS.174..571D}} In supernova-driven shocks and in the hot post-shock or coronal gas this minimum size is likely to be significantly larger\citep[{\it e.g.}, as large as particles with 100's of atoms,][]{2010A&A...510A..36M,2010A&A...510A..37M,2012A&A...545A.124B,2013A&A...556A...6B,2014A&A...570A..32B} and so DIBs would not be expected in these energetic regions. 

Based on experimental results, it would seem that if the DIB carriers are stable species they must contain $\gtrsim 40 \pm 10$ heavy X atoms because lighter species are rapidly destroyed in the ISM.\citep[{\it e.g.},][]{1994ApJ...420..307J,1996A&A...305..616A} On the other hand if they are smaller than this then the DIB carriers can apparently only be transient species, which means that their chemistry and their abundances will be sensitive to the strength and intensity of the local ISRF, particularly to  stellar UV photons. Hence it would seem that we can naturally divide the DIB carriers into two distinct classes:
\begin{enumerate}
\item large, partially dehydrogenated, radical cations resistant to UV photo-processing; \\ 
C$_n$H$_x^{p+}$ (\ $n \gtrsim 40$, \ $x \leq (2n+1)$, \ $p = 0$, 1, 2\ ), and 
\item transient small, highly dehydrogenated, radical cations sensitive to UV radiation; \\ 
C$_n$H$_x^{p+}$ (\ $n \lesssim 40$, \ $x \ll 2n$, \hspace{0.7cm} $p = 0$, 1, 2\ )
\end{enumerate} 
In the diffuse ISM it is likely that any pendant methyl, alkyl (C$_n$H$_{(2n-1)}$) or other side groups, such as those seen in the asphaltenes, would be removed through the effects of photo-dissociation/photolysis but that bridging species ought to be more resistant. In the cracked petroleum residues very large side groups are removed (but nevertheless some remain) and are therefore under-represented in the asphaltene analyses. In the following we therefore concentrate only on the core aromatic-rich carbon structures.

%------------------------------------------------------------------
\subsection{The likely chemistry of the DIB carrier (sub-)structures} 
\label{sect_likely}
%------------------------------------------------------------------ 

In terms of the possible chemical species that could be the carriers of the DIBs we would need to explore (experimentally and theoretically) many millions of possible structures (see Section~\ref{sect_riches}). As an illustration of this the asphaltene moieties in Table~\ref{table_asphaltenes} show the very rich variety and complexity of aromatic species with only $\sim 20-50$ carbon atoms (and no two are the same). However, an unguided search for viable DIB carriers (among the many millions of such large molecules) would be a formidable and inherently intractable task, almost infinitely more difficult than "looking for a needle in a haystack". This is why we have not yet been able to elucidate the fundamental nature of the DIBs carriers, with the exception of C$_{60}^+$ and its two associated DIBs.\citep{2015Natur.523..322C} The clearly difficult but fruitful experiments that eventually led to the association of C$_{60}^+$ with two DIBs was, nevertheless, aided by the fact that fullerenes have been detected in the ISM.\citep[{\it e.g.},][]{2010Sci...329.1180C} In this case the "haystack" to search was known and is rather small. Given that, other than fullerene, no large complex molecule, ion or radical has yet been identified in the ISM, the problem is in general that of finding the "haystack" before a search for the "needle" can begin. 

Given the above-discussed aromatic-rich domain characteristics gleaned from the analysis of asphaltenes we can conclude that in the ISM the aromatic-rich domains and sub-domains within a-C(:H) nano-particles and sub-micron sized grains are most likely:
\begin{itemize}
\item relatively compact, with central S$_n$ cores with $n = 4-10$, 
\item have mixed S$_n$P$_m$ structures with one or several peripheral or bridging P rings ($m \geq 1$), 
\item include P rings with common N, O, S, Si, P or B hetero-atoms and methylene bridges ($-$CH$_2-$), and  
\item are inter-connected by alkyl bridges. 
\end{itemize}
These asphaltene-type structures resemble the likely (sub-)constituents of interstellar a-C(:H) (nano-)particles\cite{2012A&A...542A..98J,2013A&A...558A..62J,2013A&A...555A..39J,2014P&SS..100...26J,2015A&A...581A..92J}, which are probably also the type of particles at the heart of fullerene formation in circumstellar regions.\citep{2012ApJ...757...41B,2012ApJ...761...35M} 

Based on the structures of the analysed PA and CA asphaltenes (Table~\ref{table_asphaltenes}) it is here suggested that an experimental and theoretical search for DIB-carrier families could be most usefully directed towards ionised and dehydrogenated $<$H]S$_n$P$^X$S$_m^+$ species with $n > 1$, $m\geq 0$ and nitrogen-doped hetero-cyclic P rings ({\it i.e.}, X = N). Here the pre-cursor $<$H] is used to indicate a dehydrogenated state. In follow up work more complex structures, such as $<$H]S$_n$P$^X$P$^X$S$_m^+$ and $<$H]S$_n$P$^X$S$_m$P$^X$S$_p^+$, where X = N, O, S, Si, P, B, Al, Ge, \ldots might also be worth exploring. 

Based on all of the above discussions, other possible DIB carrying structures could include: 
cumulene-containing aromatics or cumulene-type rings without or with hetero-atoms (principally nitrogen atoms), 
the family of couronenes, couronanes and courone-anes (also possibly doped with N).

%------------------------------------------------------------------
\subsection{DIBs and chemical associations} 
\label{sect_associations}
%------------------------------------------------------------------ 

An association between certain DIBs and small radical species, {\it e.g.}, C$_2$, C$_3$, CH, CN, OH, CH$^+$ and OH$^+$, is supported by the observational evidence but an interpretation of the correlations, anti-correlations or non-correlations is far from straight-forward. 
Nevertheless, strong DIBs do seem to trace the diffuse atomic gas\cite[{\it e.g.},][]{2014IAUS..297..153W} but only the 619.60 and 661.36\,nm DIBs appear to show a nearly perfect correlation across numerous lines of sight covering a wide range of environments.\citep{2010ApJ...708.1628M} 
Among the di-atomic radicals there does appear to be a general correlation between C$_2$, CN, CH and OH but 
these species do not seem to correlate with CH$^+$ and OH$^+$.\cite[{\it e.g.},][]{2014IAUS..297..121K} The so-called C$_2$ DIBs, which are relatively weak and seemingly associated with denser gas, also appear to correlate with C$_2$, CN and CH.\citep{2003ApJ...584..339T} 
Further, a number of studies seem to show that some DIBs correlate better with CH than with CN, 
with the C$_2$ and C$_3$ rotational temperature or with the C$_2$ column density,  
while  others anti-correlate with a high CN rotational temperature.\citep{2008A&A...484..381W,2009A&A...498..785K,2011A&A...531A..68K,2014IAUS..297..121K}
There is also evidence that some DIBs appear to be weaker towards regions where strong CN absorption is observed, {\it e.g.}, towards the supernova SN2014J in M82.\citep{2014ApJ...792..106W} 
However, in all of these works only a relatively limited number of DIBs were studied and so it is hard to generalise, particularly so because the correlations are made using a mix of observed equivalent widths and column densities. 
The correlations issue is perhaps somewhat clouded by the fact that both gas and dust  are traced by the total hydrogen column density and so will always correlate to some degree or another.
Thus, the question of DIB families is still an open one but the idea of two broad classes does appear tenable, {\it i.e.}, 1) broad and strong DIBs associated with weak molecular features and 2) weak and broad DIBs associated with the strong lines of simple radicals.\citep{2014IAUS..297...23K}
It is therefore difficult to make any firm inferences about the relationship between the DIBs, small radicals and the extinction due to interstellar nano-particles. However, and within the framework of a top-down mechanism for the formation of the DIB carriers via nano-particle fragmentation, it is perhaps possible to generally conclude that: 
\begin{enumerate} 

\item the carriers of the "C$_2$" DIBs  are seemingly associated with or formed in denser gas, 

\item the anti-correlation of some DIBs with CN could be taken to imply that nitrogen-containing hydrocarbons play a role in the DIB carrier chemistry, {\it i.e.}, when DIB carriers are destroyed nitrogen appears in the gas as CN and the DIBs are weak or absent,   

\item a weak UV bump and steep UV extinction can result from the accretion of aliphatic-rich, a-C:H mantles on grains in denser regions\citep{2012A&A...542A..98J,2013A&A...558A..62J,2015A&A...000A.000J,2015A&A...000A.000Y} but weak DIBs are not necessarily a consequence, unless the DIBs only result from aromatic-rich a-C materials, 

\item a weak UV bump and steep UV extinction is also possible in energetic regions, as a result of nano-particle destruction,\cite[{\it e.g.},][]{2010A&A...510A..36M,2014A&A...570A..32B} leading to weak or absent DIBs, and  

\item the DIBs seemingly do not correlate with small cation radicals (CH$^+$ and OH$^+$) and therefore the carriers are likely to be disfavoured in highly excited regions. 

\end{enumerate}
An observable link between nitrogen and the DIBs will probably be rather difficult to elucidate because its depletion is very low, rather invariable and it does not seem to follow that of the other elements in the transition to denser regions.\cite{2009ApJ...700.1299J} 
In the diffuse ISM the dust does not appear to be very rich in nitrogen-bearing species, for instance observations indicate that there is $\leqslant 0.3$\% of the available nitrogen in $-$C$\equiv$N bearing organics\cite{2001ApJ...550..793W} and so nitrogen can therefore only be a trace dust species, albeit a potentially very important trace element. 
Nevertheless, nitrogen does appear to be an important dopant in organic nano-globules.\cite{2006Sci...314.1439N,2008LPI....39.2391M,2009LPI....40.1130D,2009LPI....40.2260D,ANT_RSOS_globules}

%------------------------------------------------------------------
\subsection{The DIBs and the extended red emission (ERE)} 
\label{sect_H36+ERE}
%------------------------------------------------------------------ 

It is likely that the DIB carriers are related to the extended red emission (ERE) carriers and that both arise in species that are marginally-stable in a wide variety of environments, particularly in the diffuse ISM.\citep{2014IAUS..297..173W} The DIB and ERE originators must both involve carriers with electronic transitions $1.4 - 2.2$\,eV above a ground state.\cite{2014IAUS..297..173W} Interestingly, electronic transitions in the range $1.4 - 2.2$\,eV above a ground state look remarkably similar to the $1.0 - 2.6$\,eV optical gaps typical of hydrogenated amorphous carbon materials, a-C:H.\cite[{\it e.g.}, $1.0 - 2.7$\,eV][]{1986AdPhy..35..317R} As has also been pointed out the DIB and ERE carriers show non-unique global spectra, perhaps indicative of broad families of particles that are sensitive to the local environment, especially the UV radiation field.\cite{2014IAUS..297..173W} 

A particularly interesting region for ERE observations is the Red Rectangle region, which shows an unusual geometry with both internal and external (interstellar) sources of UV photons.\citep{2009ApJ...693.1946W} 
Here the ERE only occurs along the walls of an outflow cavity where there is an unobscured view of the FUV photons  from the central stars\citep{2006ApJ...653.1336V} but a blue luminescence (BL) is only seen from those regions that are shielded from the interior FUV radiation.\citep{2004ApJ...606L..65V,2005ApJ...633..262V} It appears that this geometry is rather unique and allows for a study of the UV photon-driven evolution of a-C:H grains, from wide to narrow band gap materials, and could lead to the formation of viable (pre-cursor) DIB carrier candidates.\citep{2014P&SS..100...26J,1998MNRAS.301..955D} Nevertheless, only two weak DIBs (578.0 and 661.3\,nm) have been observed in absorption in the Red Rectangle region itself or in an intervening diffuse interstellar cloud.\citep{2004ApJ...615..947H} It therefore seems seems likely that this region probably forms the DIB precursors but that the actual DIB carriers are only produced by further UV photo-processing in the low-density diffuse ISM. As was pointed out, in relation to the Red Rectangle region, the hetero-atom doping of a-C:H materials may provide a rather interesting link between the DIBs, ERE and BL.\citep{2013A&A...555A..39J,2014P&SS..100...26J} Particularly interesting is the connection between the BL, nitrogen-doped a-C:H materials and a 442.8 nm band at the  wavelength of the most prominent DIB.

Another unusual and interesting object of key relevance to our understanding of the DIBs is Hershel\,36, which is associated with a triple O star system in the M8 HII region with an elevated visible/UV stellar radiation field and a flat UV extinction ($R_{\rm V} \sim 6$). This object exhibits relatively weak DIBs, sits behind a dark lane with $A_{\rm V} \sim 4$\,mag. and $N({\rm HI}) = 8.9 \times 10^{21}$, shows (non-thermal) rotationally excited CH and CH$^+$ and vibrationally-excited H$_2$, due to strong IR radiation from an adjacent source but no CN and C$_2$.\citep{2013ApJ...773...42O,2014IAUS..297...94O,2014IAUS..297...89Y,2014IAUS..297..153W}    
Along this line of sight the DIBs are broad and many ($\sim 25\%$) show extended red wings.\citep{2013ApJ...773...42O,2014IAUS..297...89Y} This is likely another region where, as per the Red Rectangle, dense cloud carbonaceous dust is being actively UV-processed to form DIB carriers. However, unlike the  Red Rectangle, Hershel\,36 clearly shows DIBs but DIBs that are somewhat different from those observed elsewhere.

In the following section we explore a top-down fragmentation scenario for the formation of the DIB carriers.

%------------------------------------------------------------------
\section{Top-down branching and DIB carrier formation} 
\label{sect_topdown0}
%------------------------------------------------------------------ 

While no two individual interstellar nano-particles are likely to be identical the population as a whole will likely incorporate the sub-structures that actually give rise to the electronic transitions that are responsible for the DIBs. These sub-structures within nano-particles will be protected by the surrounding material and the associated transitions would likely give rise to broader DIBs that seem to more widely distributed spatially and that are relatively stable against UV.\cite[{\it e.g.},][]{2014IAUS..297..153W}  However, in a low density ISM with a hard UV radiation field the nano-particles will be partially dehydrogenated,\citep[{\it e.g.},][]{2015A&A...581A..92J} ionised and thus rendered more susceptible to dissociative fragmentation processes.\citep[{\it e.g.},][]{1996A&A...305..616A}  Their daughter fragmentation products are likely to be aromatic-rich moieties and the DIBs that these sub-nm fragments may carry would be UV-sensitive and formed in the ISM by the top-down fragmentation of hetero-atom doped a-C(:H) nano-particles. These fragments would then themselves be fragmented into non-DIB carrying small polyatomic radical species C$_2$, C$_3$, CN, CH, CH$^+$, OH, \ldots) and eventually their constituent atoms/ions (C, C$^+$, N, N$^+$, O, S, Si, P, \ldots). 
The rate of this top-down fragmentation process will be determined by both the intensity and hardness of the ambient UV radiation field and will lead to a steep sub-nano-particle fragment size distribution with a steepness that depends upon the photo-dissociation rate of all sub-nm species. This may explain why the same DIBs are sometimes seen with and without the same molecular radical species, with a much harder and/or more intense UV radiation field being present in the without case. 

% FIGURE 9 *********************************************************
\begin{figure}[!h]
 %\centering\includegraphics[width=5in]{Z_figures/branching.pdf}
 \centering\includegraphics[width=5in]{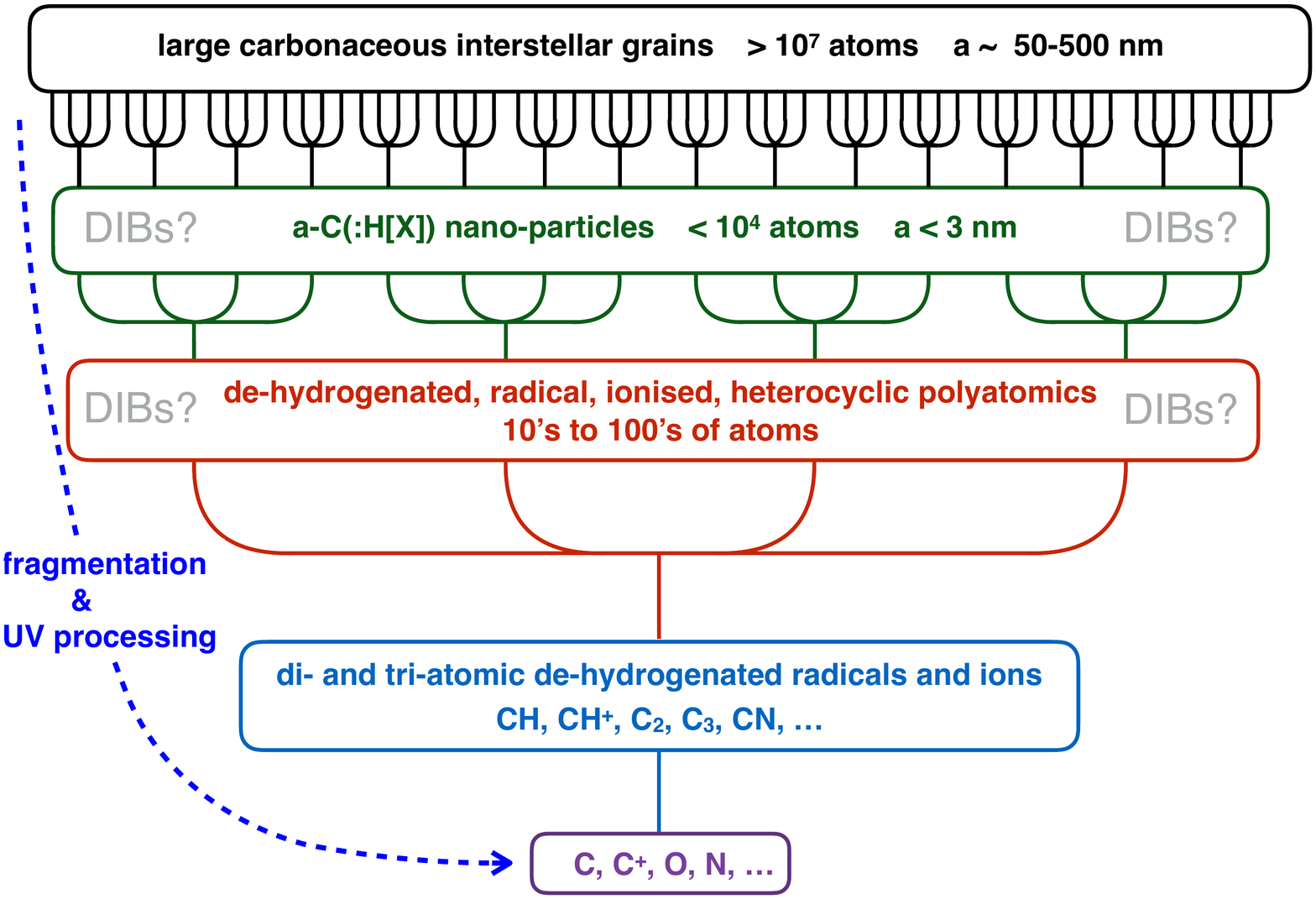}
 \caption{A schematic view of the heirarcal top-down route to the formation of the DIBs carriers by fragmentation and their formation, evolution and destruction by UV photo-processing.}
 \label{fig_branching}
\end{figure}
% *********************************************************

Here a top-down branching scenario (see Fig.~\ref{fig_branching}) is proposed for the formation of the DIB carriers. 
Top-down branching is like tree branching but the sequential evolution is from top to bottom with chemical variety decreasing from top to bottom in line with a decreasing chemical complexity and decreasing number of constituent heavy atoms. At the top of the tree are the large grains, in the middle branches the nano-particles, in the lowest branches and upper trunk can be found the molecules and at the very base of the trunk the atoms and atomic ions. 

In this scheme the number of configurations is reduced, to well below that of all possible configurations, by processes that leave only the most stable fragments. Hence, descending in size through the effects of fragmentation leads to a more limited number of smaller fragments with a more restricted conformational range.

%------------------------------------------------------------------
\subsection{Tree tops: big grains} 
\label{sect_topdown1}
%------------------------------------------------------------------ 

It has been suggested that in the ISM the big carbonaceous grains, composed principally of a-C(:H), are probably the source of all of the smaller carbon-rich grains.\citep{2012A&A...542A..98J,2013A&A...558A..62J} However, it is difficult to imagine that these grains can consist of only carbon and hydrogen atoms as they must at their sites of formation (around evolved stars and via accretion in molecular clouds) and during their sojourn in the ISM become polluted with hetero-atoms, {\it i.e.}, they are actually a-C:H:X grains doped with X hetero-atoms\citep{2013A&A...555A..39J,2014P&SS..100...26J} and/or carbonyl-rich functional groups perhaps not unlike the meteoritic and cometary organic nano-globules.\cite{ANT_RSOS_globules} This pollution must be carried down into the smaller grains in the diffuse ISM through the effects of fragmentation.\citep[{\it e.g.},][]{1996ApJ...469..740J,2012A&A...542A..98J,2013A&A...558A..62J} These polluting hetero-atoms and groups will be incorporated into the grain chemical structure resulting in interesting functional groups and structures, many of which are likely to have electronic transitions lying within the visible wavelength range.  This scenario, developed in the following section within the framework of nano-particles, would quite naturally explain why the DIBs generally correlate with the interstellar extinction, as measured by $E(B-V)$, and the dust emission, as measured by the $12\,\mu$m continuum.

It is now clear that the dust in the ISM does vary (in structure and composition) from one region to another\citep[{\it e.g.},][]{2013A&A...558A..62J,2015A&A...579A..15K,2015A&A...577A.110Y} and so a region to region environmental dependance of the DIBs is naturally to be expected. For instance, and if the DIBs are, as we propose below, due to nano-particles and smaller then there is a clear environmental explanation as to why they do not associate with the molecular hydrogen in the denser interstellar gas. This is due to the evolution of the dust properties in these regions where the dust has accreted mantles of a-C:H and has been coagulated into aggregates. Here all the small DIB-carrying nano-particles have been swept up into larger structures in which their electronic transitions are submerged by the global, aggregate dust optical properties.

% FIGURE 10 *********************************************************
\begin{figure}[!h]
 %\centering\includegraphics[width=5in]{Z_figures/coloured_heterocycles1.pdf}
 %\centering\includegraphics[width=5in]{Z_figures/coloured_heterocycles2.pdf}
 \centering\includegraphics[width=5in]{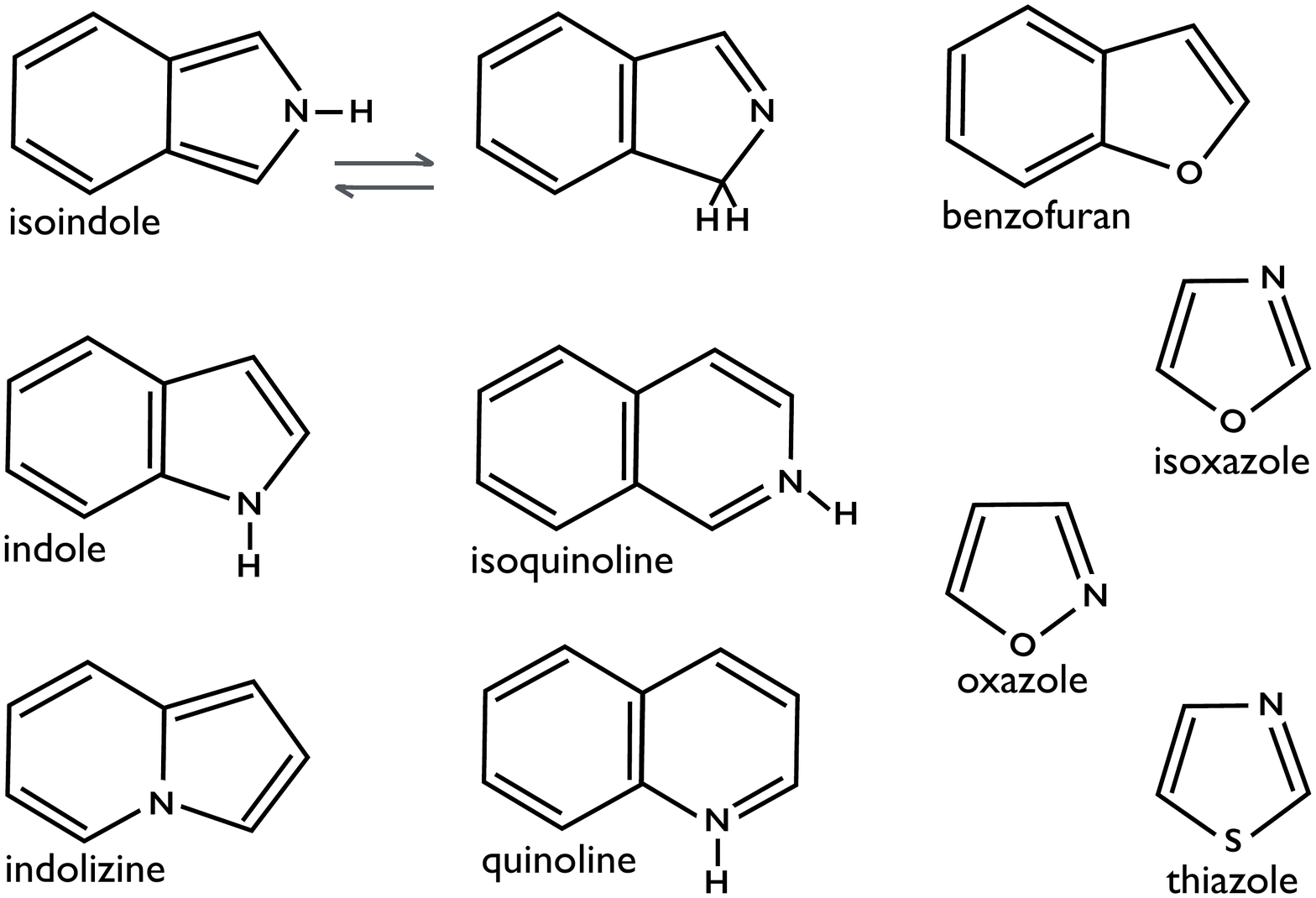}
 \centering\includegraphics[width=5in]{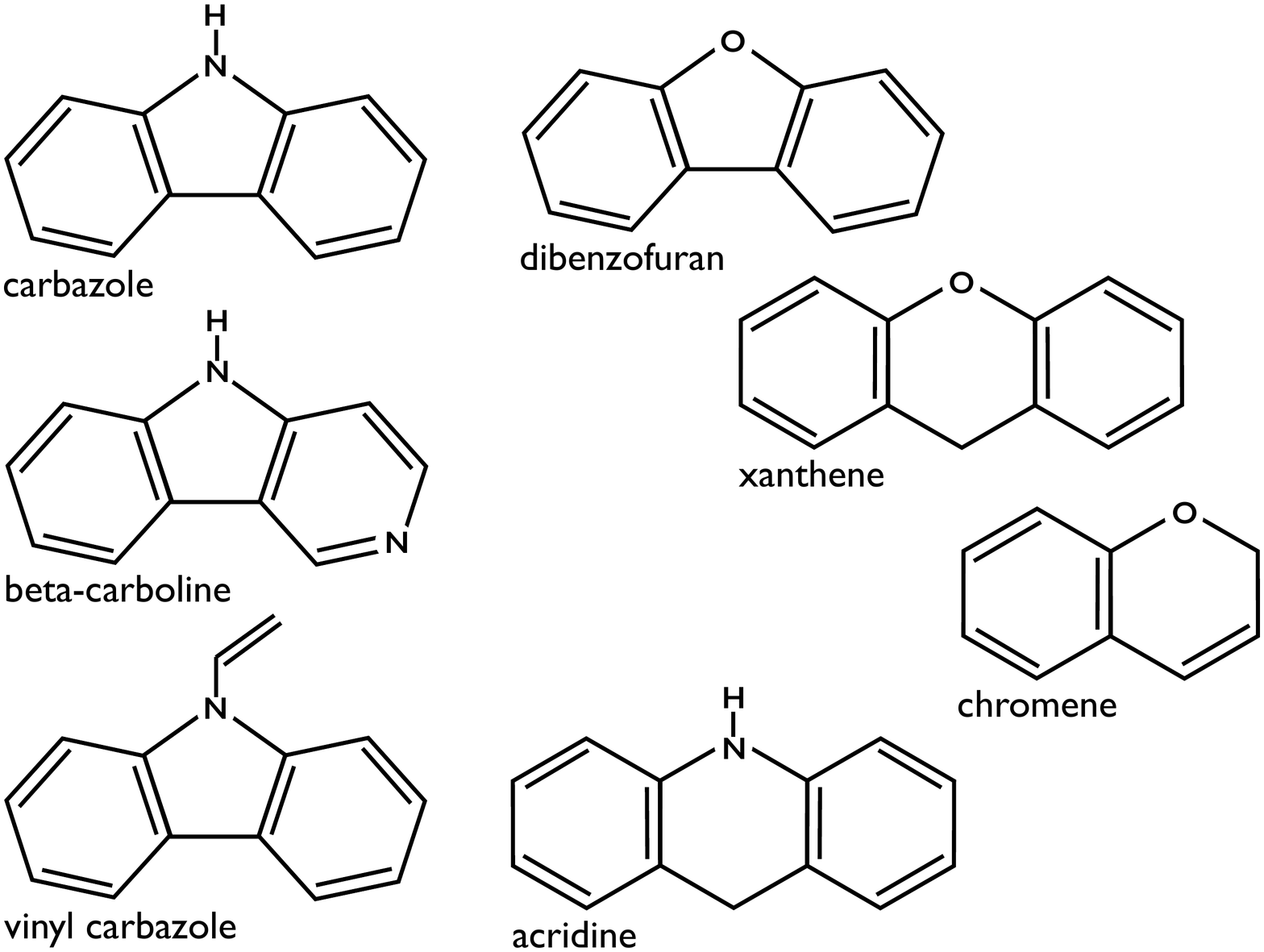}
 \caption{Heterocyclic species (P$^{\rm XX}$, SP$^{\rm X}$, S$_2^{\rm X}$, SP$^{\rm X}$S and S$_3^{\rm X}$, with X = N, O or S) with $5-15$ heavy atoms and interesting DIB colour-related properties, including that: many of them are coloured, they and their derivatives are used in dye-making and/or photo-receptors, some of them fluoresce when exposed to UV and all are stable species widespread in nature.}
 \label{fig_heterocolour}
\end{figure}
%% *********************************************************

%------------------------------------------------------------------
\subsection{Middle branches: nano-particles} 
\label{sect_topdown2}
%------------------------------------------------------------------ 

Heterocyclic sub-structures are widespread in nature in aromatic moieties and their existence in interstellar carbonaceous grains is therefore almost a certainty.\citep{2013A&A...555A..39J,2014P&SS..100...26J} Their presence in interstellar nano-particles is therefore practically unavoidable and they are likely at the heart of the more stable and UV-resistant DIBs observed in the ISM. Fig.~\ref{fig_heterocolour} shows some of the many possible heterocyclic molecular structures, including only N, O and S hetero-atoms, which are of interest to the DIB problem because many of them are coloured, used in dye-making/photo-receptors and some of them fluoresce when exposed to UV. However, heterocyclic structures with phosphorous, boron and silicon hetero-atoms are also well-known and therefore equally likely to exist within the interstellar dust population. The structures shown in Fig.~\ref{fig_heterocolour} are all well-known and distinct molecules, however, in the diffuse ISM they can probably only exist as sub-structures within larger (nano-)particles where they would be protected from the dissociating effects of the interstellar UV radiation field, {\it i.e.}, they would likely be less dehydrogenated and more neutral than ionised. These structures could perhaps therefore explain the broader DIBs that are more resistant to UV radiation DIBs. 

If interstellar nano-particles are indeed responsible for some of the DIBs then their intrinsic electronic transitions would naturally be broadened due to the perturbations introduced by their immediate environment within the larger grain structure. Thus, the responsible electronic transitions are intrinsically broader than any atomic or molecular lines. 

Given that the heterocyclic structures likely to be at the heart of the DIBs will undergo a top-down processing, structurally-related DIBs will form one from another. For example, in Fig.~\ref{fig_heterocolour} it can be seen that adding or removing rings from some of the structures leads to their inter-conversion, {\it e.g.}, the molecules $\beta$-carboline or carbazole are so related to the smaller isoindole and pyrrole molecules (the latter a five-fold ring with one nitrogen hetero-atom is not shown) and similarly dibenzofuran and benzofuran, and also xanthene and chromene. Hence, it is not unexpected that inter-DIB correlations do not always go through zero because the formation of some DIB structures will probably depend upon the destruction or transformation of others. Further,  that the DIBs generally do not correlate well with each other is probably due to the wide variety of possible structures that can give rise to electronic transitions at the wavelengths of interest for the DIBs. 

No significant correlation between the UV bump, the far-UV extinction rise and the DIB strengths has been found.\cite[{\it e.g.},][]{2011IAUS..280..162C}   
However, in the Small Magellanic Cloud it appears that when the UV bump is weak or absent then so are the DIBs.\cite{2014IAUS..297..147C} 
Thus, there is some evidence that when the interstellar nano-particles responsible for the UV bump are destroyed then so are the DIB carriers and, as noted above (Section 6 subsection \ref{sect_associations}), the anti-correlation of some DIBs with CN seems to indiate that nitrogen may plays a role in the DIB carrier chemistry. Hence, by inference, it would appear that nitrogen could be an important hetero-atom dopant (at the \% level) in interstellar hydrocarbon nano-particles.

%------------------------------------------------------------------
\subsection{Low branches: radical and molecular ions} 
\label{sect_topdown3}
%------------------------------------------------------------------ 

% FIGURE 11 *********************************************************
\begin{figure}[!h]
 %\centering\includegraphics[width=5in]{Z_figures/DIB_hyd.pdf}
 %\centering\includegraphics[width=5in]{Z_figures/DIB_dehyd.pdf}
 \centering\includegraphics[width=5in]{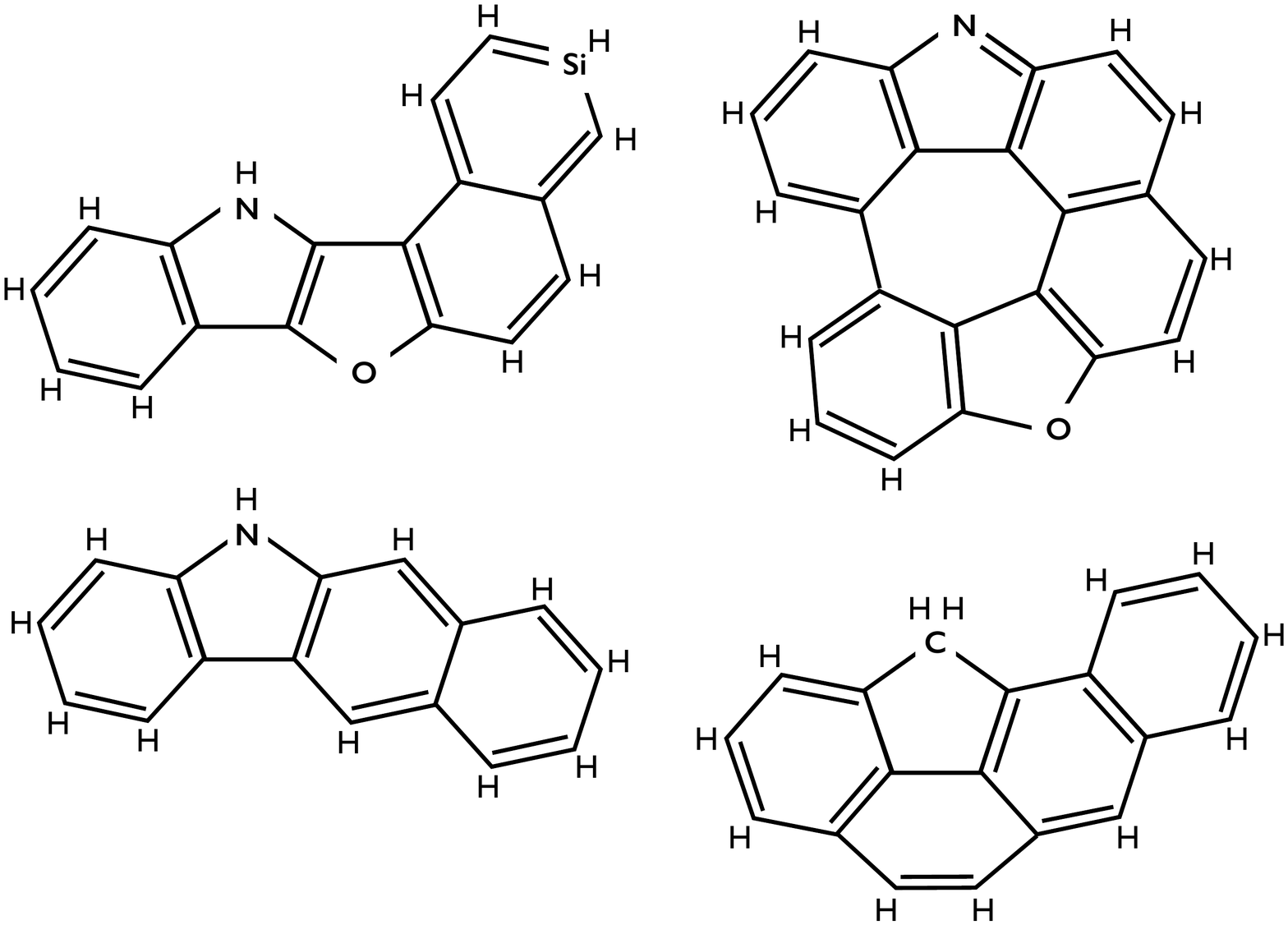}
 \centering\includegraphics[width=5in]{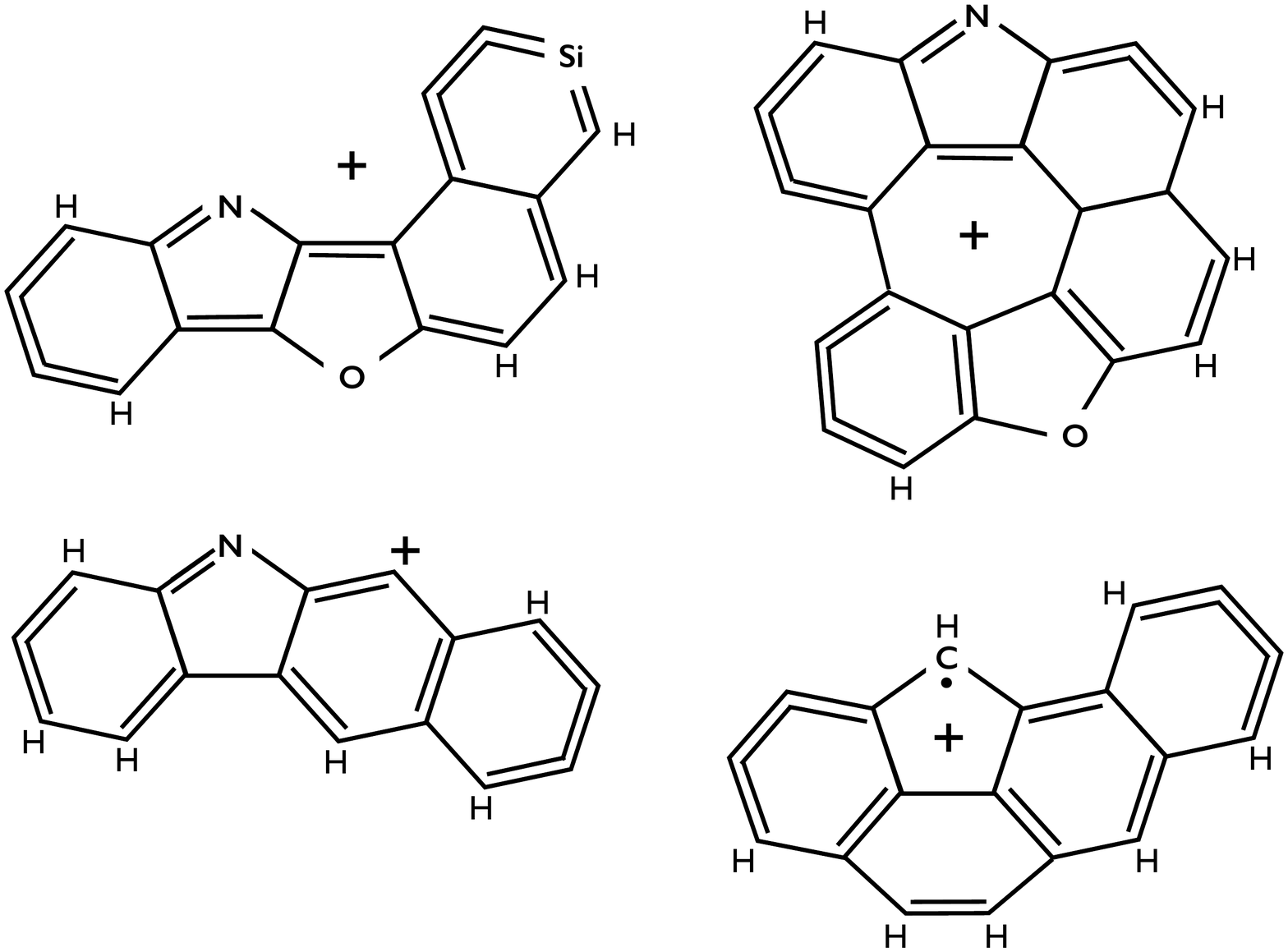}
 \caption{Four likely DIB pre-cursor molecules, with $17-24$ heavy atoms and the following structures S$_2$P$^{\rm O}$P$^{\rm N}$S, S$_2$P$^{\rm O}$SGSP$^{\rm N}$, S$_2$P$^{\rm N}$S and S$_4$P$^\prime$. The lower part of the figure shows their ionised and $\approx 50$\% dehydrogenated forms, 
 $<$H]S$_2$P$^{\rm O}$P$^{\rm N}$S$^+$, 
 $<$H]S$_2$P$^{\rm O}$SGSP$^{\rm N}$$^+$, 
 $<$H]S$_2$P$^{\rm N}$S$^+$ and 
 $<$H]S$_4$P$^\prime$$^+$: 
 such polycyclic, hetero-cyclic, radical cations could be candidate DIB carriers.}
 \label{fig_DIBdehyd}
\end{figure}
% *********************************************************

The disruption of interstellar nano-particles, most likely driven by UV photon processing (through photo-dissociation and charge effects), will yield smaller fragments that are then even less stable in the hard UV radiation field than are their parent grains. In this case the liberated species will now look more like molecules than nano-particles but must obviously suffer the consequences and bear the marks of dehydrogenation and exist as cations in these harsh environments.\cite[{\it e.g.},][]{2011IAUS..280..162C}  Nevertheless, the observed structure in DIBs that appear to be somewhat UV-resistant does seem to be consistent with the rotational contours of rather $5-40$ atom gas phase molecules.\cite{1976MNRAS.174..571D,2013ApJ...773...42O}

\begin{figure}[!h]
 %\centering\includegraphics[width=5in]{Z_figures/top-down.pdf}
 \centering\includegraphics[width=5in]{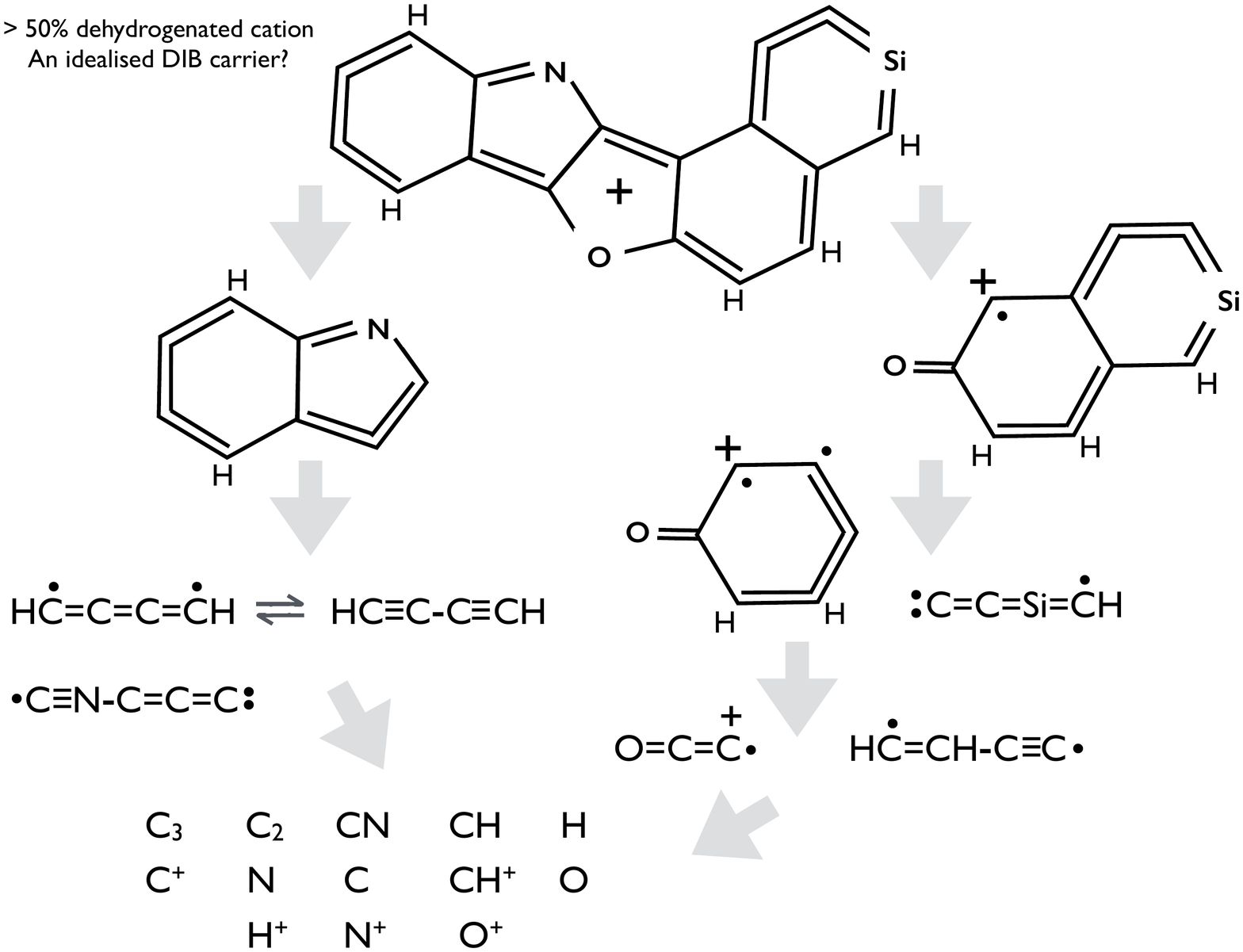}
 \caption{A schematic scenario for a top-down evolution of the DIB carriers. Note that the proposed  DIB carrier, 
 $<$H]S$_2^{\rm Si}$P$^{\rm O}$P$^{\rm N}$S$^+$, is shown with unrealistically abundant hetero-atoms (N, O and Si) in order to illustrate the possible evolutionary pathways that will need to be explored in more detail.}
 \label{fig_top-down}
\end{figure}

The upper four hetero-atomic, polycyclic species in Fig.~\ref{fig_DIBdehyd} (C$_{17}$H$_{11}$NOSi, C$_{22}$H$_{10}$NO,  C$_{16}$H$_{11}$N and C$_{19}$H$_{12}$, clockwise from top left) with S$_2$P$^{\rm O}$P$^{\rm N}$S, S$_2$P$^{\rm O}$SGP$^{\rm N}$S, S$_2$P$^{\rm O}$P$^{\rm N}$S and S$_4$P$^\prime$ structures are some possible DIB pre-cursor species, which are shown as rather hetero-atom rich ({\it e.g.}, N, O and Si containing) for illustrative purposes only. These structures are all viable, heterocyclic aromatic moieties, having less than 30 heavy atoms, that would not be stable against UV photo-dissociation in the diffuse ISM. The likely DIB-carrying forms of these species, their $\approx 50$\% dehydrogenated cations, are shown in the lower part of the figure. These types of aromatic, radical, cation, moieties will also be dissociated by UV photons in the diffuse ISM but could represent the sort of top-down transient fragments that therefore carry the UV-sensitive DIBs. A schematic scenario for the likely top-down evolution of the DIB carriers is shown in Fig.~\ref{fig_top-down}, where the proposed DIB carrier is illustratively shown with N, O and Si hetero-atoms, corresponding to about a 14\% (3/21) heavy atom doping level. 

% FIGURE 12 *********************************************************
% *********************************************************

Given the discussion in the previous sub-section, it is most probable that nitrogen with its intrinsic propensity to incorporate into aromatic rings, and especially five-fold rings, will be the most important aromatic cycle hetero-atom.  Nitrogen readily inserts into these structures and is known to form a myriad of interesting organic molecules, such as pyrroles, pyridines, pyrimidines, azoles, indoles, quinolines, {\it etc.} However, dihetero-atomic cycles could perhaps play a more important role because it appears that many of them exhibit many colourful properties, especially species with two nitrogen atoms ({\it e.g.}, imidazole, pyrimidine) or a nitrogen and an oxygen atom ({\it e.g.}, oxazoles).

%------------------------------------------------------------------
\subsection{The trunk: di- and tri-atomic radical ions, \ldots} 
\label{sect_topdown4}
%------------------------------------------------------------------ 

The photo-dissociation of the radical and molecular ions discussed in the above sub-section would quite naturally lead to the formation of di- and tri-atomic radical ions such as CH, CH$^+$, CN, C$_2$ and C$_3$, which appear to bear some relation with the observed DIBs. For instance, the observation of the same DIBs with or without these same radicals in different regions of the ISM could be explained by variations in the local UV radiation field, which would determine the rate of photo-dissociation and therefore how quickly the nano-particle fragments are photo-dissociated into smaller radical species, {\it e.g.}, C$_2$, C$_3$ and CN, which are then themselves photo-dissociated into their constituent atoms and ions C, C$^+$, N, {\it etc.} 

In particular, weaker DIBs appear to be associated with strong CN absorption,\citep[{\it e.g.},][]{2014ApJ...792..106W} which would be natural if some of the DIBs are due to the type of nitrogen-containing moieties proposed above and whose end-of-the-line photo-dissociation/fragmentation products would then naturally include CN. Also, the observation of CN with a high rotational temperature in an interstellar cloud that appears to exhibit no DIBs \citep{2011A&A...531A..68K} could be explained by the very rapid destruction of the DIB carriers to their excited di-atomic  and atomic/ionised constituents. Once again supporting the view that nitrogen is an important (hetero-atomic) element for the DIBs.

%------------------------------------------------------------------
\subsection{The base: the atomic/ionised ISM} 
\label{sect_topdown5}
%------------------------------------------------------------------ 

At the base of the tree, and in the most diffuse ISM where molecular radicals are rapidly UV photo-dissociated, we find the isolated atoms and ions in the gas, principally, H, C, C$^+$, O, N, Si$^+$ and S$^+$, some of which were constituents within the dust there. The ultimate product of UV photo-processing are the small polyatomic radicals and ions such as CH, CH$^+$, CN, C$_2$ and C$_3$.
Further, in regions of the ISM where there has been heavy UV processing and where significant dust destruction, especially of carbon-rich nano-particles, has occurred then it appears there are no DIBs, {\it i.e.}, there can be no DIBs without nano-particles but there probably can be nano-particles without DIBs.

%------------------------------------------------------------------
\section{Experimental and observational considerations}
\label{sect_experiment}
%------------------------------------------------------------------

Given that an experimental exploration of the many millions of possible DIB carriers is an intractable problem it would be best to take a statistical approach and therefore explore numerous possibilities at the same time. Such an approach will allow a considerable narrowing of the search once DIB(-like) transitions have been found in a viable material. With this kind of approach in mind the following suggestions may prove useful in determining a long-term experimental programme with the aim of finding viable classes of materials that could be, or hopefully are, the DIB carriers, {\it i.e.}, with the aim of finding the haystacks within which to search for the needles. 

It is here suggested that a fruitful experimental search should therefore be directed towards exploring the: 
\begin{enumerate}
\item Low temperature NIR-UV spectroscopy of asphaltenes both with and without sample irradiation by FUV photons, which would drive both critical sub-structure dehydrogenation and cation formation.
\item Low temperature NIR-UV spectroscopy of a series of single hetero-atom (X = N, O, S, Si, P, \ldots) doped a-C(:H):X materials both with and without FUV irradiation (as per 1.). Initially, at least, the aromatic-rich and aliphatic-rich near-end members of the a-a-C(:H):X material family should be explored, with a later exploration of the middle ground materials. 
\item Follow-up spectroscopy under identical conditions (as per 1. and 2. above) but for mixed hetero-atom (N+O, N+S, N+Si, N+P, {\it etc.}) doped a-C(:H):X:Y and eventually triply-doped (N+O+S, N+O+Si, N+O+P, \ldots) a-C(:H):X:Y:Z materials.
\end{enumerate}
Once DIB-carrying solid phases have been determined the relevant carriers should then be subjected to closer scrutiny to locate the the sub-structures responsible for the DIBs. Further, this approach could then be combined with the low-temperature visible/near-IR spectroscopy of colour-carrying aromatic moieties, such as those shown in Fig.~\ref{fig_heterocolour}, both with and without FUV-irradiation in order to explore the neutral and dehydrogenated/ionised forms. 

It also appears that there might be observational avenues that could yet yield something interesting. For example, the relationship of the DIBs with line-of-sight depletions, particularly with that of nitrogen, does not appear to have been extensively studied and this could perhaps be worth a deeper exploration.

%------------------------------------------------------------------
\section{Conclusions}
\label{sect_conclusions}
%------------------------------------------------------------------

It seems that taking a global view of interstellar dust and gas evolution, within the framework of a top-down fragmentation model, may provide a promising new way to view the nature of the DIBs. The diffuse ISM is indeed a harsh environment and small species, with less than many tens of heavy atoms, do not do well there. Thus, the UV processing of interstellar matter at nm and sub-nm size scales would appear to link together and lie at the heart of many of the dust and gas variations that are observed in space.  

The origin of the DIBs almost certainly lies in the heterocyclic five- and six-fold ring aromatic moieties that form an integral part of the  contiguous structure of interstellar nm and sub-nm sized particles, {\it i.e.}, they are most likely specific, multi-ring (aromatic), sub-structures within nano-particles. In particular, the aromatic sub-structures within the nano-particles will be somewhat protected from the effects of interstellar UV radiation and could explain the more UV-resistant DIB population. However, once these moieties are released from nano-particles as "free-flying" species they will be significantly dehydrogenated and will also be ionised to their cationic forms. Such species in the gas phase will be significantly more susceptible to UV photolysis than their neutral and fully hydrogenated counterparts once liberated from their parent nano-particles. They could thus explain the UV-sensitive DIB population. Thus, it would appear that some significant fraction of the DIB carriers could be "unstable", transient species that are rather sensitive to the local physical conditions. 

Significantly dehydrogenated and ionised aromatic moieties are much less stable than their parent hydrogenated neutrals in the diffuse ISM and will not necessarily re-form the original aromatic (PAH)  structure upon re-hydrogenation by atomic H from the gas phase. They may rather fragment to form smaller ring systems and/or linear chains. Thus, for many highly dehydrogenated aromatic structures (<H]PAHs) in the ISM there may be no way back home to their parental structures upon reaction with gas phase atoms and ions.

The ideas presented here do not necessarily simplify the experimental or observational search for the exact form of the DIB carriers, for if they were rather simple we would have long ago found the solution to this long-standing astro-chemical-physical problem. 

The author hopes that the ideas presented here might help to focus laboratory and theoretical studies along a track that may, on the way to the desired destination, uncover some interesting science and perhaps even elucidate the generic nature of the undoubtedly complex species that are the carriers of the diffuse interstellar bands.

%%%%%%%%%%%%%%%%%%%%%%%%%%%%%%%%%%%%%%%%%
\section*{Acknowledgment} 
The author would like to thank everyone he has ever talked to about interstellar dust but, in particular, special thanks to Dan Welty and Don York for valuable discussions and guidance on the interpretation of DIB observations. 

%%%%%%%%%%%%%%%%%%%%%%%%%%%%%%%%%%%%%%%%%

%% for the bibliography, at the end
\footnotesize{
\bibliography{Ant_bibliography} % your references Yourfile.bib
\bibliographystyle{rsc} 
}

\end{document}